\documentclass[twocolumn,pre,floats,superscriptaddress]{revtex4-1}

\newcommand{\tw}{\ensuremath{t_\mathrm{w}}\xspace}

\newcommand{\Tg}{\ensuremath{T_\mathrm{g}}\xspace}
\newcommand{\Tm}{\ensuremath{T_\mathrm{m}}\xspace}

\usepackage[T1]{fontenc}
\usepackage[utf8]{inputenc}
\usepackage{graphicx}
\usepackage{amsmath}
\usepackage{amssymb}
\usepackage{color}
\usepackage{xspace}
\usepackage{multirow}
\usepackage{hyperref}
\usepackage{dcolumn}
\makeatletter
\newcommand*{\balancecolsandclearpage}{%
  \close@column@grid
  \cleardoublepage
  \twocolumngrid
}
\makeatother
\graphicspath{{./figures/}{./}}

\begin{document}

\title{Scaling law describes the spin-glass response in theory, experiments and simulations}

\author{Q. Zhai} \email{These authors contributed equally to 
  this work.} \affiliation{Texas
  Materials Institute, The University of Texas at Austin, Austin,
  Texas 78712, USA}
\author{I.~Paga} \email{These authors contributed equally to 
  this work.} \affiliation{Dipartimento di Fisica, Sapienza Universit\`a di Roma,
  INFN, Sezione di Roma 1,Italy} \affiliation{Departamento de
  F\'\i{}sica Te\'orica, Universidad Complutense, 28040 Madrid, Spain}

\author{M.~Baity-Jesi}\affiliation{Eawag,  Überlandstrasse 133, CH-8600 Dübendorf, Switzerland}

\author{E.~Calore}\affiliation{Dipartimento di Fisica e Scienze della
  Terra, Universit\`a di Ferrara e INFN, Sezione di Ferrara, I-44122
  Ferrara, Italy}

\author{A.~Cruz}\affiliation{Departamento de F\'\i{}sica Te\'orica,
  Universidad de Zaragoza, 50009 Zaragoza,
  Spain}\affiliation{Instituto de Biocomputaci\'on y F\'{\i}sica de
  Sistemas Complejos (BIFI), 50018 Zaragoza, Spain}

\author{L.A.~Fernandez}\affiliation{Departamento de F\'\i{}sica
  Te\'orica, Universidad Complutense, 28040 Madrid,
  Spain}\affiliation{Instituto de Biocomputaci\'on y F\'{\i}sica de
  Sistemas Complejos (BIFI), 50018 Zaragoza, Spain}

\author{J.M.~Gil-Narvion}\affiliation{Instituto de Biocomputaci\'on y
  F\'{\i}sica de Sistemas Complejos (BIFI), 50018 Zaragoza, Spain}

\author{I.~Gonzalez-Adalid Pemartin}\affiliation{Departamento  de F\'\i{}sica Te\'orica, Universidad Complutense, 28040 Madrid, Spain}

\author{A.~Gordillo-Guerrero}\affiliation{Departamento de
  Ingenier\'{\i}a El\'ectrica, Electr\'onica y Autom\'atica, U. de
  Extremadura, 10003, C\'aceres, Spain}\affiliation{Instituto de
  Computaci\'on Cient\'{\i}fica Avanzada (ICCAEx), Universidad de
  Extremadura, 06006 Badajoz, Spain}\affiliation{Instituto de
  Biocomputaci\'on y F\'{\i}sica de Sistemas Complejos (BIFI), 50018
  Zaragoza, Spain}

\author{D.~I\~niguez}\affiliation{Instituto de Biocomputaci\'on y
  F\'{\i}sica de Sistemas Complejos (BIFI), 50018 Zaragoza,
  Spain}\affiliation{Fundaci\'on ARAID, Diputaci\'on General de
  Arag\'on, Zaragoza, Spain}

\author{A.~Maiorano}\affiliation{Dipartimento di Biotecnologie, Chimica e
  Farmacia, Università degli studi di Siena, 53100, Siena,
  Italy}\affiliation{INFN, Sezione di Roma 1, I-00185 Rome,
  Italy}\affiliation{Instituto de Biocomputaci\'on y F\'{\i}sica de Sistemas
Complejos (BIFI), 50018 Zaragoza, Spain}

\author{E.~Marinari}\affiliation{Dipartimento di Fisica, Sapienza
  Universit\`a di Roma, and CNR-Nanotec,
  I-00185 Rome, Italy}\affiliation{INFN, Sezione di Roma 1, I-00185 Rome,
  Italy}

\author{V.~Martin-Mayor}\affiliation{Departamento de F\'\i{}sica
  Te\'orica, Universidad Complutense, 28040 Madrid,
  Spain}\affiliation{Instituto de Biocomputaci\'on y F\'{\i}sica de
  Sistemas Complejos (BIFI), 50018 Zaragoza, Spain}

\author{J.~Moreno-Gordo}\affiliation{Instituto de Biocomputaci\'on y
  F\'{\i}sica de Sistemas Complejos (BIFI), 50018 Zaragoza,
  Spain}\affiliation{Departamento de F\'\i{}sica Te\'orica,
  Universidad de Zaragoza, 50009 Zaragoza, Spain}

\author{A.~Mu\~noz-Sudupe}\affiliation{Departamento de F\'\i{}sica
  Te\'orica, Universidad Complutense, 28040 Madrid,
  Spain}\affiliation{Instituto de Biocomputaci\'on y F\'{\i}sica de
  Sistemas Complejos (BIFI), 50018 Zaragoza, Spain}

\author{D.~Navarro}\affiliation{Departamento de Ingenier\'{\i}a,
  Electr\'onica y Comunicaciones and I3A, U. de Zaragoza, 50018
  Zaragoza, Spain}

\author{R.~L.~Orbach}
\affiliation{Texas Materials Institute, The University of Texas at Austin,
  Austin, Texas  78712, USA}

\author{G.~Parisi}\affiliation{Dipartimento di Fisica, Sapienza
  Universit\`a di Roma, INFN, and CNR-Nanotec,
  I-00185 Rome, Italy}\affiliation{INFN, Sezione di Roma 1, I-00185 Rome,
  Italy}

\author{S.~Perez-Gaviro}\affiliation{Escuela Universitaria Politécnica - La
  Almunia, 50100 La Almunia de Doña Godina, Zaragoza,
  Spain}\affiliation{Instituto de Biocomputaci\'on y F\'{\i}sica de Sistemas
  Complejos (BIFI), 50018 Zaragoza, Spain}\affiliation{Departamento de
  F\'\i{}sica Te\'orica, Universidad de Zaragoza, 50009 Zaragoza, Spain}

\author{F.~Ricci-Tersenghi}\affiliation{Dipartimento di Fisica, Sapienza
  Universit\`a di Roma, and CNR-Nanotec,
  I-00185 Rome, Italy}\affiliation{INFN, Sezione di Roma 1, I-00185 Rome,
  Italy}

\author{J.J.~Ruiz-Lorenzo}\affiliation{Departamento de F\'{\i}sica,
  Universidad de Extremadura, 06006 Badajoz,
  Spain}\affiliation{Instituto de Computaci\'on Cient\'{\i}fica
  Avanzada (ICCAEx), Universidad de Extremadura, 06006 Badajoz,
  Spain}\affiliation{Instituto de Biocomputaci\'on y F\'{\i}sica de
  Sistemas Complejos (BIFI), 50018 Zaragoza, Spain}

\author{S.F.~Schifano}\affiliation{Dipartimento di Scienze Chimiche e Farmaceutiche, Università di Ferrara e INFN  Sezione di Ferrara, I-44122 Ferrara, Italy}

\author{D.~ L.~ Schlagel} \affiliation{Division of Materials
  Science and Engineering, Ames Laboratory, Ames, Iowa 50011, USA}

\author{B.~Seoane}\affiliation{Departamento de F\'\i{}sica
  Te\'orica, Universidad Complutense, 28040 Madrid,
  Spain}\affiliation{Instituto de Biocomputaci\'on y F\'{\i}sica de
  Sistemas Complejos (BIFI), 50018 Zaragoza, Spain}

\author{A.~Tarancon}\affiliation{Departamento de F\'\i{}sica
  Te\'orica, Universidad de Zaragoza, 50009 Zaragoza,
  Spain}\affiliation{Instituto de Biocomputaci\'on y F\'{\i}sica de
  Sistemas Complejos (BIFI), 50018 Zaragoza, Spain}

\author{R.~Tripiccione}\affiliation{Dipartimento di Fisica e Scienze
  della Terra, Universit\`a di Ferrara e INFN, Sezione di Ferrara,
  I-44122 Ferrara, Italy}

\author{D.~Yllanes}\email{david.yllanes@czbiohub.org}\affiliation{Chan Zuckerberg Biohub, San Francisco, CA, 94158}
\affiliation{Instituto de Biocomputaci\'on y F\'{\i}sica de
  Sistemas Complejos (BIFI), 50018 Zaragoza, Spain}


\date{\today}

\begin{abstract}
The correlation length $\xi$, a key quantity in glassy dynamics, can now be
precisely measured for spin glasses both in experiments and in simulations.
However, known analysis methods lead to discrepancies either for large
external fields or close to the glass temperature.  We solve this problem by
introducing a scaling law that takes into account both the magnetic field
and the time-dependent spin-glass correlation length. The scaling law is
successfully tested against experimental measurements in a CuMn single
crystal and against large-scale simulations on the Janus~II dedicated
computer.
\end{abstract}

\maketitle 

The dynamical arrest found upon cooling glass formers (spin glasses, fragile
molecular glasses, polymers, colloids, etc.) to their glass temperature $\Tg$
is a major open problem ~\cite{cavagna:09,charbonneau:14}. In the longstanding description~\cite{adam:65}, 
this slowing down is caused by the unbounded expansion of cooperative regions  
as \Tg is approached or as the system is left to age below \Tg, which, in turn,
leads to growing free-energy barriers. A quantitative description of this 
process is usually attempted in terms of a correlation length $\xi$. Unfortunately, 
in numerical simulations it is extremely difficult to measure the 
quantities that are easily accessible to experiments (and vice versa), which has 
led to seemingly irreconcilable approaches to the computation of 
the correlation length. On the one hand, theorists study correlation
functions in an abstract replica
space~\cite{marinari:96,janus:08b,janus:09b,janus:10b,manssen:15,manssen:15b,fernandez:15,lulli:15,janus:17b,janus:18,fernandez:19b}.
On the other hand, experimentalists measure the system's response to an
applied external field (either an electric field for glass-forming
liquids~\cite{albert:16} or a magnetic field for spin
glasses~\cite{joh:99,bert:04,guchhait:14,guchhait:17,zhai:19}). Ref.~\cite{janus:17b}
reconciled the two approaches by measuring the experimental response functions
in a numerical simulation, but it was ultimately based on an  approximate
scaling law that breaks down for large fields or close to the glass temperature $\Tg$.
This is especially problematic, since  temperatures $T\approx\Tg$ are the most
relevant for the study of glass formers ($\xi$ is restricted to a very narrow window of variation
if we move away from \Tg).

Here we are able to solve this dilemma in a framework that completely harmonizes experiments
with theory. We conduct a parallel study of non-equilibrium spin-glass dynamics both in an
experiment in a CuMn single crystal
and in a large-scale simulation of the Ising-Edwards-Anderson (IEA) model carried
out on the Janus~II custom-built supercomputer~\cite{janus:14}. We introduce
a scaling law that describes the system's response over its entire natural range 
of variation.

\begin{figure}[t]
\centering \includegraphics[width =\columnwidth]{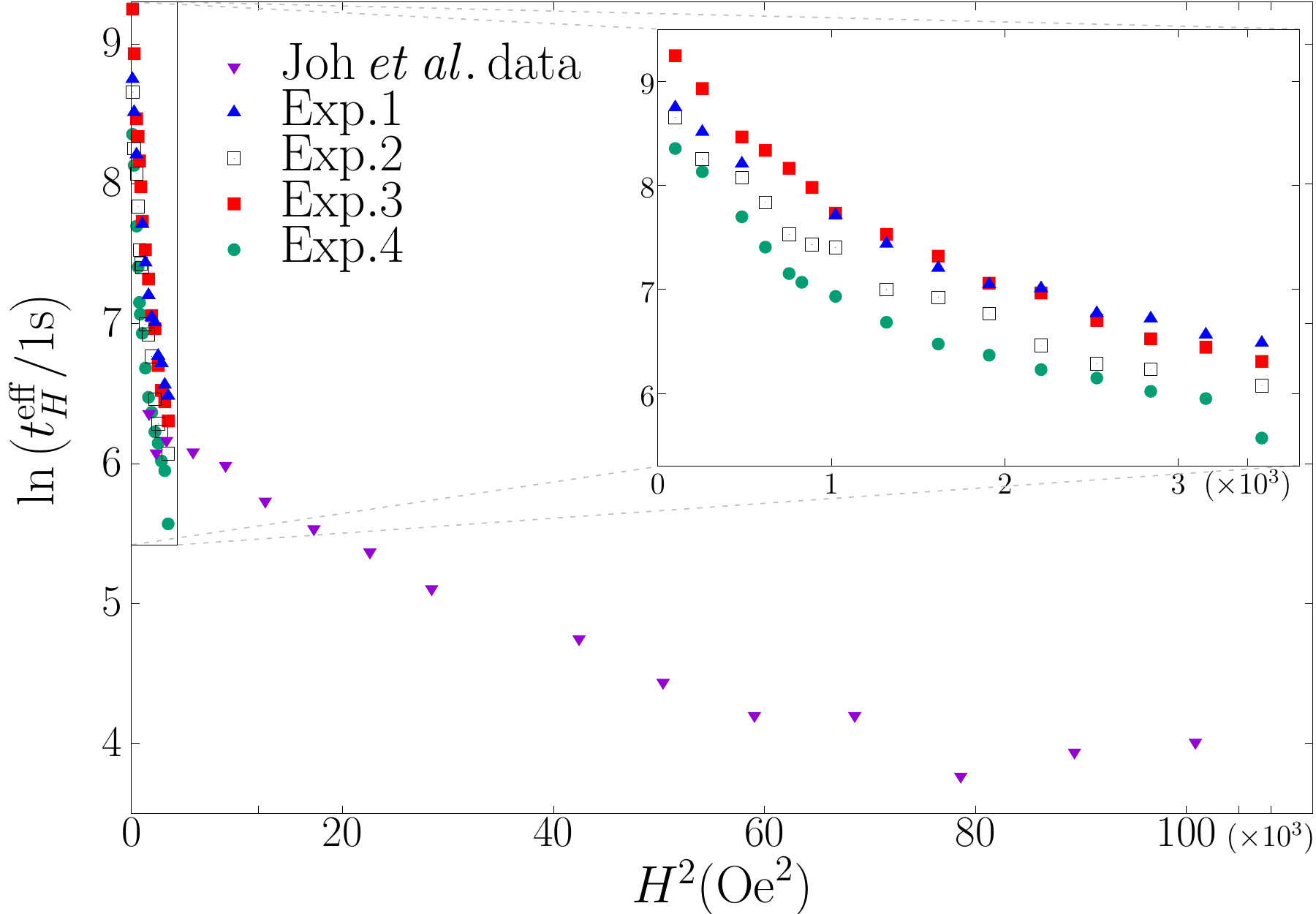}
\caption{{ Comparison of the experimental situation in 1999 (data from Joh
    et al.~\cite{joh:99}) and 2020 (present work)}. Both data sets are for
  Cu$_{94}$Mn$_6$. The 1999 data are from a poly-crystalline sample, while
  2020 data come from a single crystal allowing for a much larger correlation
  length $\xi$ (see Table~\ref{tab:Xi} for details).  The figure shows the
  maximum of the relaxation function as a function of the squared magnetic
  field $H^2$. It is easy to estimate the slope at $H^2=0$ (from which $\xi$
  is measured) for the 1999 data, which display a linear behavior for
  $H^2\lesssim 6\times 10^4$ Oe$^2$. Instead, the large $\xi$ of the 2020 data
  not only causes a larger slope, but also a much larger curvature (see the
  zoomed in region in the {\bf inset}) which makes it challenging to extrapolate
  the slope to $H^2=0$.}
\label{fig:joh99}
\end{figure}

To be specific, let us consider the zero-field-cooled protocol (see,
e.g.,~\cite{zhai:19}), where the spin glass is suddenly quenched from a
temperature well above $\Tg$ down to the working temperature $\Tm<\Tg$ and is
then left to relax for a time $\tw$ [the growth of the correlation length
$\xi(\tw)$ is unbounded for $T<\Tg$, but very slow].  At time
$\tw$, a magnetic field $H$ is applied and the growing magnetization
$M(t,\tw;H)$ is recorded for times $t+\tw$ (the $\tw$ dependence is included
because spin glasses perennially age at $T<\Tg$, slowly approaching
equilibrium but never reaching it). The maximum of the relaxation function
$\mathrm{d} (M/H)/\mathrm{d}\ln t$ defines a time $t^{\text{eff}}_H$ directly
related to the height of the free-energy barriers that the system
encounters. In a magnetic field, the Zeeman effect lowers these barriers by an
amount proportional to $H^2$ and to the number of spins in a glassy
cluster. Therefore, an Arrhenius law would predict a linear behavior of
$\ln t^{\text{eff}}_H$ with $H^2$. Yet, see Fig.~\ref{fig:joh99}, departures
from a straight line were observed for large values of $H^2$ in the very first
experiment using this approach~\cite{joh:99}.  In fact, the Zeeman
interpretation has been disputed~\cite{vincent:95,bert:04} and identifying a
linear behavior in $H^2$ becomes problematic close to $\Tg$~\cite{zhai:19}.

\begin{figure}[t]
	\centering
	\includegraphics[width=\columnwidth]{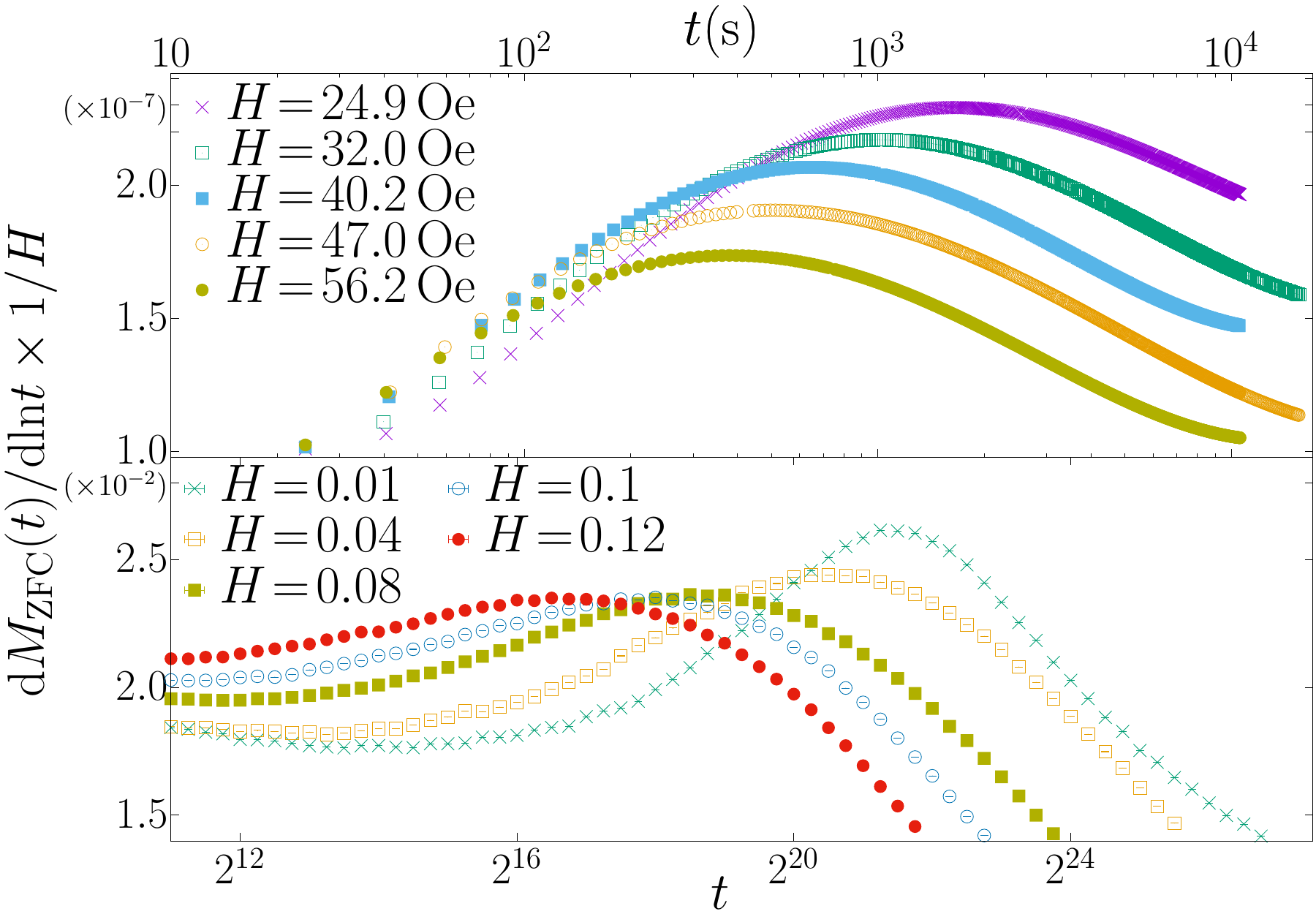}
	\caption{A set of relaxation curves
          $S(t)= \mathrm{d} (M/H)/\mathrm{d}\ln t$ for CuMn at $T=29$~K and
          $\tw=10^4$~s (\textbf{top}) and for the Ising-Edwards-Anderson model at
          $T=0.9$ and $\tw=2^{22}$ lattice sweeps (\textbf{bottom}). The relation
          between IEA and physical units is discussed in the text.           
	\label{fig:S_of_t}}
\end{figure}

In what follows, we shall derive a scaling law for the response to the
magnetic field that is still valid for large fields and close to $\Tg$.
As we stated above, the scaling law is tested against measurements in a single CuMn crystal and
against massive numerical simulations carried out on Janus~II. The single
crystal is important because the growth of $\xi(\tw)$
is not limited like in a poly-crystal with grain boundaries~\cite{kenning:18}.
Specifically, we shall show that the $H$ dependence
has the form
\begin{equation}\label{eq:prediction}
\ln \frac{t^{\text{eff}}_H}{t^{\text{eff}}_{H\to 0^+}}=
\frac{\hat S}{T} \xi^{D-\frac{\theta}{2}} H^2\ +\ \xi^{-\theta/2} {\cal G}\big(\xi^{D-\frac{\theta}{2}}H^2;T\big)\,.
\end{equation}
Here $\xi$ stands for $\xi(\tw)$, $\hat S$ is
a constant, $D=3$ is the spatial dimension and $\theta$ stands for the 
replicon exponent $\theta(\tilde x)$~\cite{marinari:96,janus:08b,janus:09b}, 
where $\tilde{x}=\ell_{\mathrm{J}}(T)/\xi(t_\mathrm{w})$
and $\ell_\mathrm{J}(T)$ is the Josephson
length~\cite{janus:18,zhai:19}.  

For small values of $x$ the scaling function behaves as ${\cal G}(x)\sim x^2$
($x=\xi^{D-\frac{\theta}{2}}H^2$).  Hence, ${\cal G}$ is of order $H^4$ for
small values of the magnetic field and, if $\xi$ is small (the typical case
well below $\Tg$), the contribution of ${\cal G}$ can be neglected for small
$H$. In fact, most previous experiments and simulations only tested the $H^2$
term in Eq.~\eqref{eq:prediction}. We find here, however, that for larger
fields, or larger correlation lengths (which are found only close to $\Tg$),
${\cal G}$ is the dominant contribution. Fortunately, Eq.~\eqref{eq:prediction}
offers a unified framework that rationalizes the entire range of experiment and
simulations.

\paragraph*{Experimental and  numerical descriptions.}
Our experiments used
a commercial DC SQUID to measure the magnetization of a Cu$_{94}$Mn$_6$
single crystal with $\Tg=31.5$~K, grown at Ames Laboratory, U.S.
DOE (see~\cite{zhai:19} for details). The sample was
quenched from 40 K at 10 K/min to the measuring temperature $\Tm$ in zero
magnetic field.  After the temperature was stabilized, the system was aged for
a waiting time $\tw$ before a magnetic field $H$ was turned on, and the
magnetization $M_{\text {ZFC}}(t,\tw;\Tm)$ was recorded as a function of time $t$.
The temperatures were chosen as 28.5~K, 28.75~K and 29~K, so
$\Tm\ge 0.9 \Tg$. The magnetic fields ranged from 16~Oe to 59~Oe.
Table~\ref{tab:Xi} shows the relevant experimental parameters, 
including the effective replicon exponent $\theta(\tilde x)$.
\begin{table}
\begin{ruledtabular}
	\begin{tabular}{c  c c    c    c    c}
		& & \Tm (K) & $\tw$ (s) & $\xi(\tw)/a$ & $\theta(\tilde x)$ \\
                \hline
		\textbf{Exp. 1} && 28.50 & 10\,000 &320.36 &0.337\\
		\textbf{Exp. 2} && 28.75 & 10\,000 &341.76 &0.344\\
		\textbf{Exp. 3} && 28.75 & 20\,000 &359.18 &0.342\\
		\textbf{Exp. 4} && 29.00 & 10\,000 &391.27 &0.349\\
	\end{tabular}
\end{ruledtabular}
        \caption{Main parameters for our four experiments, including
          the correlation length at time $\tw$ (in units of the average Mn-Mn spacing
          $a$) and the effective replicon exponent
          $\theta(\tilde x)$, obtained from the interpolation in~\cite{zhai:19} of
          the results in~\cite{janus:18}.}
        	\label{tab:Xi}
\end{table}

In parallel with these experiments, we have simulated the Ising-Edwards-Anderson (IEA) model, with Hamiltonian
$\mathcal H = -\sum_{\langle \boldsymbol x, \boldsymbol y\rangle}
J_{\boldsymbol x \boldsymbol y} s_{\boldsymbol x}s_{\boldsymbol y} - H
\sum_{\boldsymbol x} s_{\boldsymbol x}$, where $s_{\boldsymbol x}=\pm1$ is the
spin at site $\boldsymbol x$.  We have used one sample of a cubic lattice
with periodic boundary conditions, linear size $L=160$ and random
couplings $J_{\boldsymbol x\boldsymbol y}=\pm 1$~\footnote{In a $L=160$
  system, $\xi(\tw)$ and $M(t,\tw;H)$ display little sample dependence,
  see~\cite{zhai-janus:20b}. We have, however, run 512 independent
  thermal histories for our sample (the benefits of simulating many
  independent thermal histories are discussed in Ref.~\cite{janus:18})}.  In
these natural units, and for $H=0$, the IEA model undergoes a spin-glass phase
transition at the critical temperature $\Tg=1.102(3)$ \cite{janus:13}. We
simulated the non-equilibrium dynamics by means of a Metropolis algorithm. The
natural time unit is the lattice sweep, which roughly corresponds to one
picosecond of physical time. As for the magnetic field, Ref.~\cite{janus:17b}
estimated from experimental Fe$_{0.5}$Mn$_{0.5}$TiO$_3$
data~\cite{aruga_katori:94} that $H=1$ in the IEA model corresponds to
$5\times 10^4$~Oe.

In order to mimic the experimental setup in the simulations, an initial random
spin configuration is placed instantaneously at the working temperature \Tm
and left to relax for a time $\tw$, with $H=0$. At time $\tw$, the external
magnetic field is turned on and  the magnetization
$M(t,\tw;H)$ and the correlation function
$C(t,\tw;H)=\sum_{\boldsymbol x}\ s_{\boldsymbol x}(\tw;H=0)\, s_{\boldsymbol x}
(t+\tw;H)/160^3\,$ are recorded.

Our experimental range ($16$~Oe to $59$~Oe)
corresponds to $0.0003\lesssim H \lesssim 0.0012$ in the IEA model, but the signal-to-noise
ratio limited our simulations to $H\geq 0.005$. 
We employed two tricks to
match these scales.
On the one hand, we can use dimensional analysis~\cite{fisher:85} 
to relate $H$ and the reduced temperature $\hat t=(\Tg-T)/\Tg$ through
\begin{equation}\label{eq:fisher-sompolinsky-suggestion}
{\hat t}_{\mathrm{num}}\approx {\hat t}_{\mathrm{exp}} \,\left(
\frac{H_{\mathrm{num}}}{H_{\mathrm{exp}}} \right)^{\frac{4}{\nu(5-\eta)}}\,,
\end{equation}
where $\nu=2.56(4)$ and $\eta=-0.390(4)$ are $H=0$ critical
exponents~\cite{janus:13}, while the subscripts exp and num
stand for experiment and simulation.  Eq.~\eqref{eq:fisher-sompolinsky-suggestion} suggests 
that we increase ${\hat t}_{\mathrm{num}}$ to reach the experimental
scale with our range or $H_\text{num}$, which results in
$0.89\lesssim T_\mathrm{num}\lesssim 0.99$. Given our pre-existing database
of long simulations at $H=0$~\cite{janus:18}, it has been convenient
to work at temperatures $\Tm=0.9$ and $\Tm=1.0$ (or $\hat t=0.183$ and $0.093$).

On the other hand, we have found
that, when $H\to 0$, the correlation function $C(t,\tw;H)$ approaches a
constant value $C_{\text{peak}}$ at the maximum of the relaxation
function~\cite{zhai-janus:20b}, which suggests computing $t^{\text{eff}}_H$
in the simulations from the equation
\begin{equation}\label{eq:C_peak}
  C(t^{\text{eff}}_H,\tw;H)=C_{\text{peak}}\,.
\end{equation}
See~\cite{gonzalez-adalid-pemartin:19} for a similar choice in an equilibrium
context. This is helpful because Eq.~\eqref{eq:C_peak} can be solved at $H=0$
as well [in contrast with the magnetization, $C(t,\tw;H)$ does not vanish at
$H=0$].  The values of $C_\text{peak}$ are given in Table~\ref{tab:Cpeak}.

	\begin{table}
\begin{ruledtabular}
		\begin{tabular}{c c  c l D{.}{.}{-1} c c} 
			& & \Tm &\multicolumn{1}{c}{$\tw$} & \multicolumn{1}{c}{$\xi(\tw,H=0)$} & $\theta(\tilde x)$& $C_\text{peak}$ \\ [0.5ex] 
			\hline		
			\textbf{Run 1} & & 0.9 & $2^{22}$     & 8.294(7)  & 0.455  & 0.530  \\ 
			\textbf{Run 2} & & 0.9 & $2^{26.5}$   & 11.72(2)  & 0.436  & 0.510  \\ 
			\textbf{Run 3} & & 0.9 & $2^{31.25}$  & 16.63(5)  & 0.415  & 0.490  \\
			\textbf{Run 4} & & 1.0 & $2^{23.75}$  & 11.79(2)  & 0.512  & 0.419  \\
			\textbf{Run 5} & & 1.0 & $2^{27.625}$ & 16.56(5)  & 0.498  & 0.400  \\
			\textbf{Run 6} & & 1.0 & $2^{31.75}$  & 23.63(14) & 0.484  & 0.383  \\ [1ex] 
		\end{tabular}
\end{ruledtabular}

		\caption{Main parameters for our numerical simulations,
                  including the replicon exponent $\theta(\tilde x)$
		   and the value of $C_\mathrm{peak}$ employed in
                  Eq.~\eqref{eq:C_peak}.}
		\label{tab:Cpeak}
	\end{table}

\paragraph*{The scaling law.}
        
We work here on the same assumptions of Ref.~\cite{janus:17b},
though we shall be able to improve on their findings.

In equilibrium and for large-enough correlation lengths, a scaling
theory describes  the magnetic response to an external field $H$
\cite{parisi:88,amit:05}.  Our assumption will be (see also
Refs.~\cite{fernandez:15,lulli:15}) that this scaling theory holds as well in
the non-equilibrium regime, at least for large $\xi(\tw)$ and small $H$:
\begin{equation}
\begin{split}
M(t,\tw;H) &= [\xi(t+\tw)]^{-\frac{D}{2}-\frac{\theta(\tilde{x})}{4}}\\ & \times {\cal F} \left( H [ \xi
  ( t + \tw) ] ^{\frac{D}{2}-\frac{\theta(\tilde{x})}{4}} , \frac{\xi(t +\tw)}{\xi(\tw)} \right)\,,
\label{eq:m-scaling}
\end{split}
\end{equation}
Because of (at least approximate) full-aging spin-glass dynamics (see,
e.g.,~\cite{rodriguez:03}), Eq.~\eqref{eq:C_peak} tells us that
$\xi(t +\tw)/\xi(\tw)$ will be approximately constant close to the maximum of
the relaxation rate (see Fig.~\ref{fig:S_of_t}), and we shall omit this
dependence. Taylor expanding
Eq.~\eqref{eq:m-scaling}, and recalling that ${\cal F}(x)=-{\cal F}(-x)$, we find
 \begin{equation}\label{eq:suscept-defined}
M(t,\tw;H)=\chi_1 H+ \frac{\chi_3}{3!} H^3+ \frac{\chi_5}{5!} H^5+\mathcal{O}(H^7)\,,
\end{equation}
where~\footnote{In order to not overburden the notation we have omitted the $t$ and
$\tw$ arguments in the r.h.s. susceptibilities in
Eq.~\eqref{eq:suscept-defined}.} 
\begin{equation}
\label{eq:chi-scaling}
\chi_{2n-1} \propto b_{2n}(T)[\xi(\tw)]^{(n-1)D-\frac{n\theta(\tilde{x})}{2}}\,
\end{equation}
[$b_{2n}(T)$ is a smooth function of $T$].

Our improvements over the results of~\cite{janus:17b} start from the
observation that Eq.~\eqref{eq:chi-scaling} predicts the paradoxical result
$\chi_1\propto \xi^{-\theta(\tilde{x})/2}$ (hence, $\chi_1$ would go to zero when
$\xi\to\infty$). In fact, Eq.~\eqref{eq:chi-scaling} neglects the contribution
of the regular part of the free energy. A better description, then, is
\begin{equation}\label{eq:corrections-chi1}
\chi_1=\frac{\hat S(C_{\text{peak}})}{T}+ \frac{b_2(T)}{\xi^{\theta(\tilde{x})/2}}\;,
\end{equation}
where $\hat S\bigl(C(t,\tw)\bigr)$ is the function appearing in the
fluctuation-dissipation relations
(FDR)~\cite{cugliandolo:93,marinari:98f,franz:98,janus:16} [from now on, we
use the shorthand $\hat S$ for $\hat S(C_{\text{peak}})$].

Our next assumption will be that we can determine the excess free energy per
spin in a field as it is done in equilibrium (by integrating
$M$ with respect to $H$)
\begin{equation}
\label{eq:free_energy_vs_chi_expansion}
\Delta F =-\left[\frac{\chi_1}{2} H^2+ \frac{\chi_3}{4!} H^4+ \frac{\chi_5}{6!} H^6+\mathcal{O}(H^8)\right]\,.
\end{equation}
Eq.~\eqref{eq:free_energy_vs_chi_expansion}, combined with
Eqs.~(\ref{eq:chi-scaling},\ref{eq:corrections-chi1}) leads directly to
Eq.~\eqref{eq:prediction} when one makes a few additional
hypothesis~\footnote{For the sake of simplicity, we have neglected the
  correction of order $\xi^{-\theta/2}$ in the $H^2$ term in
  Eq.~\eqref{eq:prediction}.}: (i) according to an Arrhenius law,
see~\cite{joh:99,vincent:95,djurberg:95},
$t^{\text{eff}}_H/t^{\text{eff}}_{H=0}=\text{exp}[N \Delta F/(k_\mathrm{B}
T)]$ where $N$ is the number of spins in a glassy domain, and (ii)
$N\propto \xi^{D-\theta(\tilde{x})/2}$~\cite{janus:17b}.

The prefactor $\xi^{-\theta(\tilde{x})/2}$ for the
${\cal G}$ term in Eq.~\eqref{eq:prediction}, not included in Ref.~\cite{janus:17b},
will be crucial here because, unlike
in~\cite{janus:17b}, we shall test Eq.~\eqref{eq:prediction} in
situations where the ${\cal G}$ term is the dominant contribution.

\begin{figure}
	\centering
	\includegraphics[width = 1\columnwidth]{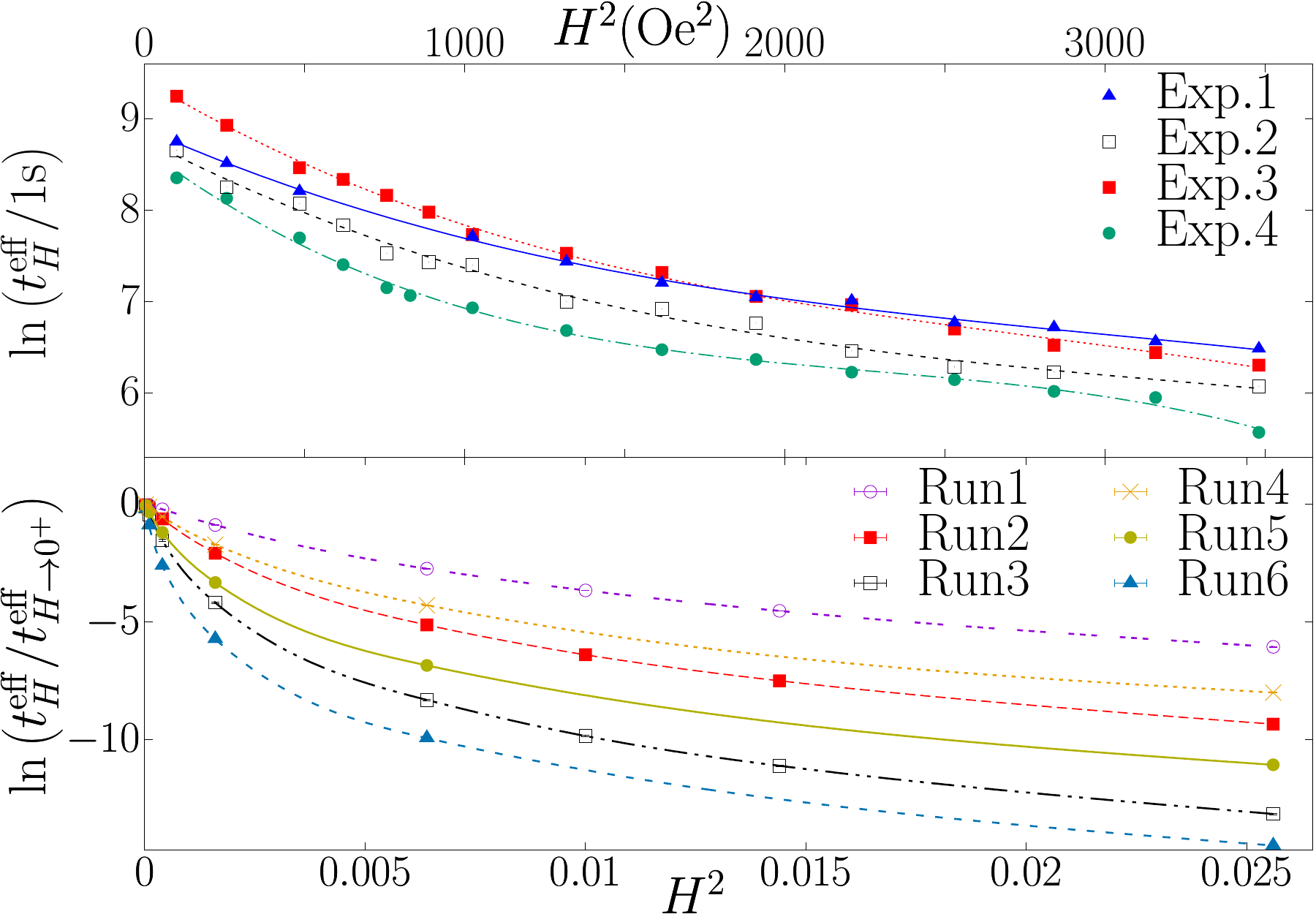}
	\caption{Experimental and 
          numerical  $\ln \tw^{\text{eff}}$ from the
          maximum of the response function in Fig.~\ref{fig:S_of_t}.
           \textbf{Top}: Data from the experiments in Table~\ref{tab:Xi}.
           Lines are fits to a polynomial in $H^2$, as in
          Eq.~\eqref{eq:coeficiente-an-def}. The  fit parameters  are
          reported in Table~\ref{tb:raw}.
          \textbf{Bottom}: Numerical data for the runs
          in Table~\ref{tab:Cpeak} (the lines are just
          guides for the eye). 
\label{fig:ratio_time_vs_H2}
}
\end{figure}

\paragraph*{Experimental and numerical results.}
We look at relaxation function curves exhibited in Fig.~\ref{fig:S_of_t},
from which the
effective times $t^{\text{eff}}_H$ are obtained. Our results for
$\ln t^{\text{eff}}_H$ (experiment) and
$\ln t^{\text{eff}}_H/t^{\text{eff}}_{H=0}$ (simulations) are displayed in
Fig.~\ref{fig:ratio_time_vs_H2}. The technical details about this analysis
will appear elsewhere~\cite{zhai-janus:20b}.  Both the experimental and the
numerical data in Fig.~\ref{fig:ratio_time_vs_H2} deviate very significantly
from linear behavior, which suggests that the ${\cal G}$ term in
Eq.~\eqref{eq:prediction} is, indeed, playing a dominant role.

Our next step is fitting the experimental data to
\begin{equation}
  \label{eq:coeficiente-an-def}
\ln t^{\text{eff}}_H = a_0 + a_2 H^2 +  a_4 H^4 + a_6 H^6 + {\cal O}(H^8) \,.
\end{equation}
Note that, in the experiments, $\ln t^{\text{eff}}_{H}$ needs to be
extrapolated to $H=0$ (this is the meaning of the $a_0$ term). Our
coefficients $a_n$ are listed in Table~\ref{tb:raw}. We extract $\xi$ from the $a_2$
term as explained in Ref.~\cite{zhai:19}.  For the higher-order terms,
Eqs.~(\ref{eq:prediction},\ref{eq:suscept-defined}) predict
$a_n\propto b_{2n}(T)\xi^{[nD-(n+1)\theta(\tilde{x})]/2}$. For instance, the $T=28.75$~K data with
$t_{\mathrm{w}}^{(1)} = 10$~ks and $\tw^{(2)} = 20$~ks allow a direct test of the
scaling relation. Taking for $\theta(\tilde{x})$ the average value
$\theta = 0.343$~\cite{zhai-janus:20b} we find
\begin{equation}
\begin{aligned}
&\xi(t_{\mathrm{w}}^{(2)})/\xi(t_{\mathrm{w}}^{(1)}) = 
\left[a_2(t_{\mathrm{w}}^{(2)})/a_2(t_{\mathrm{w}}^{(1)})\right]^{\frac{1}{D-\theta/2}} \ = 1.053,\\
&\xi(t_{\mathrm{w}}^{(2)})/\xi(t_{\mathrm{w}}^{(1)}) = 
\left[a_4(t_{\mathrm{w}}^{(2)})/a_4(t_{\mathrm{w}}^{(1)})\right]^{\frac{1}{2D-\frac{3\theta}{2}}} = 1.048, \\
&\xi(t_{\mathrm{w}}^{(2)})/\xi(t_{\mathrm{w}}^{(1)}) = 
\left[a_6(t_{\mathrm{w}}^{(2)})/a_6(t_{\mathrm{w}}^{(1)})\right]^{\frac{1}{3D-2\theta}} = 1.052.
\end{aligned}
\end{equation}

We can, therefore, gain access to the ${\cal G}$ term in Eq.~\eqref{eq:prediction}
by subtracting $a_0+a_2H^2$ from the experimental
value of $\ln t^{\text{eff}}_H$. 

As for the numerical data, polynomial fits analogous to
Eq.~\eqref{eq:coeficiente-an-def} are possible, but result in wildly
oscillating curves. The simplest explanation for this behavior is that our
largest magnetic fields are beyond the radius of convergence of the Taylor
expansion of Eq.~\eqref{eq:prediction}. One can, however,
compute $a_2$ by estimating the
derivative of $\ln (t^{\text{eff}}_H/t^{\text{eff}}_{H \to 0^+})$ numerically at
$H^2=0$~\cite{zhai-janus:20b}. Hence, we can access the ${\cal G}$
term in
Eq.~\eqref{eq:prediction} with the same subtraction that we used for the
experimental data. 

\begin{table}[t]
\begin{ruledtabular}
	\begin{tabular}{c  c  c  r c l}
		\Tm (K) & $\tw$(s) & coefficient & \multicolumn{3}{c}{value}\\
                \hline
		\multirow{3}{2em} {28.5}& \multirow {3}{2em}{10000}& $a_2$ &  $-1.551 \times 10^{-3}$  &$\pm$& $1.03 \times 10^{-4}$\\
		& &$a_4$ & $3.980 \times 10^{-7}$  &$\pm$&$6.99 \times 10^{-8}$\\
		& &$a_6$ & $-4.363\times 10^{-11}$  &$\pm$&$1.29 \times 10^{-11}$\\
		\hline
		\multirow{3}{2em} {28.75}& \multirow{3}{2em}{10000}& $a_2$ &  $-1.816 \times 10^{-3}$  &$\pm$& $2.00 \times 10^{-4}$\\
		& &$a_4$ & $4.565 \times 10^{-7}$  &$\pm$&$1.32 \times 10^{-7}$\\
		& &$a_6$ & $-4.584\times 10^{-11}$  &$\pm$&$2.45\times 10^{-11}$\\
		\hline
		\multirow{3}{2em} {28.75}& \multirow{3}{2em}{20000}& $a_2$ &  $-2.104 \times 10^{-3}$  &$\pm$& $1.19 \times 10^{-4}$\\
		& &$a_4$ & $5.889 \times 10^{-7}$  &$\pm$&$7.88 \times 10^{-8}$\\
		& &$a_6$ & $-7.013\times 10^{-11}$  &$\pm$&$1.47\times 10^{-11}$\\
		\hline
		\multirow{3}{2em} {29}& \multirow{3}{2em}{10000}& $a_2$ &  $-2.609 \times 10^{-3}$  &$\pm$& $1.28 \times 10^{-4}$\\
		& &$a_4$ & $1.016 \times 10^{-6}$  &$\pm$& $8.45\times 10^{-8}$\\
		& &$a_6$ & $-1.491\times 10^{-10}$  &$\pm$&$1.57\times 10^{-11}$\\
	\end{tabular} 
\end{ruledtabular}
	\caption{Experimental data: coefficients $a_n$ of the
          polynomial fit of $\ln \tw^\text{eff}$, see
          Eq. \eqref{eq:coeficiente-an-def}, as a function of \Tm
          and $\tw$.}
	\label{tb:raw}
\end{table}

\begin{figure}[h]
	\centering
	\includegraphics[width = 1\columnwidth]{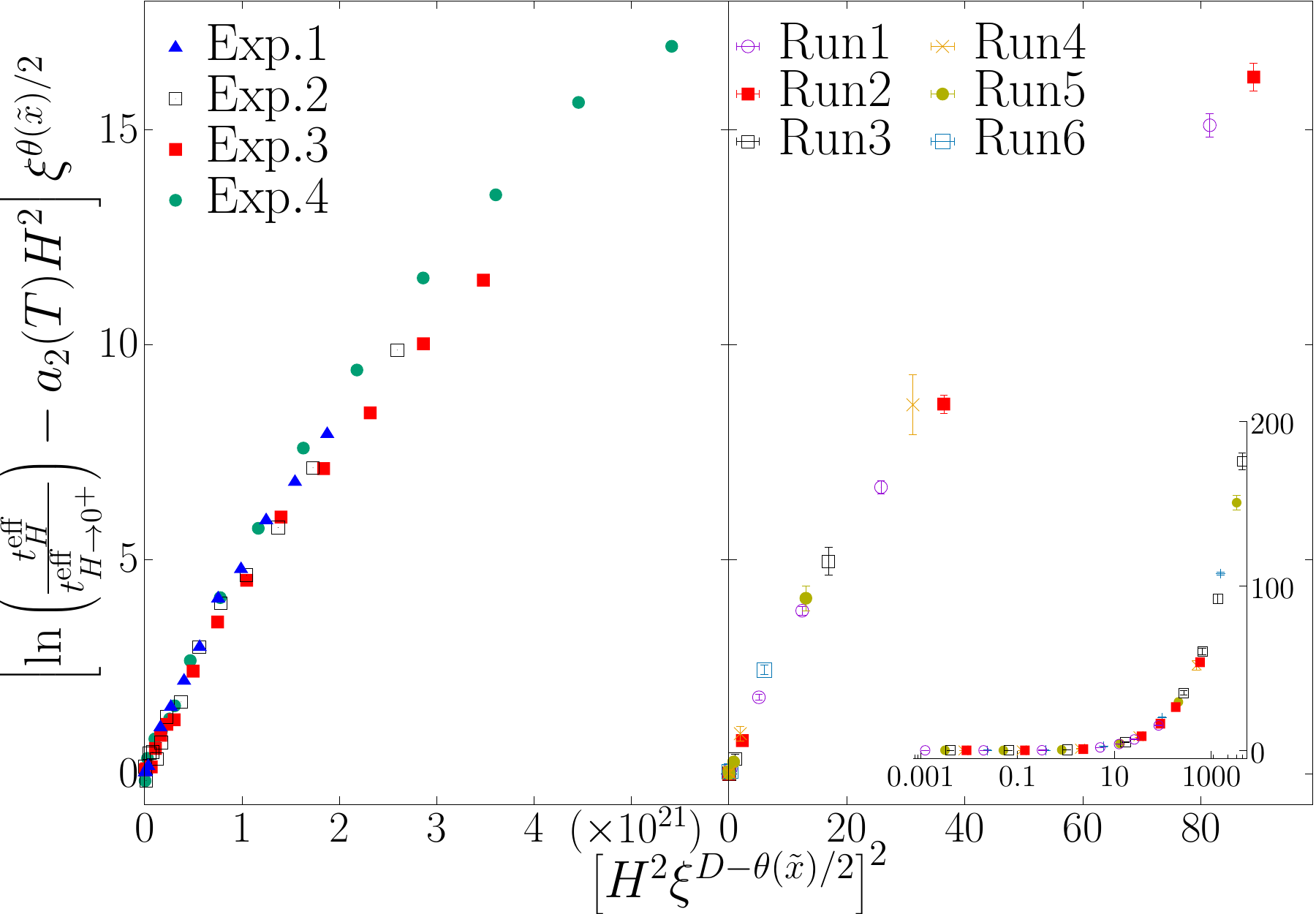}
	\caption{\label{fig:scaling_law} The non-linear part of the
          response time data: $[\ln (
          t_H^{\text{eff}}/t^{\text{eff}}_{H \to 0^+})-a_2(T)
          H^2]\xi^{\theta(\tilde{x})/2}$ plotted against the scaling variable
          $[H^2\xi^{D-\theta(\tilde{x})/2}]^2$, see Eq.~\eqref{eq:prediction}.
          \textbf{Left}: Experimental
          data (see Table \ref{tab:Xi}).
          \textbf{Right}: Numerical data (see Table \ref{tab:Cpeak}). The
          \textbf{main panel}, in
          linear scale, shows a closeup for small values of
          $[H^2\xi^{D-\theta(\tilde{x})/2}]^2$. The \textbf{inset} is in
          log scale  in order to report all our numerical data. Note that
           experimental and numerical data are reported in different 
          unit systems (see main text).}
      \end{figure}

Finally, Fig.~\ref{fig:scaling_law} brings these analyses together to perform a
strong test of Eq.~\eqref{eq:prediction} (assuming that the coefficients $b_4$
and $b_6$ are almost constant in the temperature range of interest).  The
agreement with the scaling prediction, manifested in a data collapse, is
striking both for the experimental and the numerical data~\footnote{Note that
the scaling form of Eq.~\eqref{eq:prediction} is \emph{necessary} to obtain a
good collapse. To support this, we show in the Appendix
that the collapse of Fig.~\ref{fig:scaling_law}
deteriorates if one assumes different scaling behaviors.}.

\paragraph*{Conclusions.}
The melding of experiment, theory and simulations, as exhibited in
Figs.~(\ref{fig:S_of_t})--(\ref{fig:scaling_law}), is a spectacular success of
statistical mechanics. If the right questions are asked, a truly schematic
model (namely the Ising-Edwards-Anderson model) turns out to 
behave, quantitatively, in the same way that CuMn does. The crucial ingredients to uncover this
universal behavior have been high-quality simulations carried out on a
custom-built computer, careful experiments capable of addressing the relevant regime
of very large correlation lengths close to the glass temperature, and an
extension to the non-equilibrium context of the classical equilibrium scaling theory.
We are now able to model quantitatively ---in a framework that encompasses both
experiments and numerical simulations--- responses, autocorrelation lengths, and
energy barriers in three-dimensional spin glasses. This will allow us to
address more exotic phenomena such as rejuvenation (temperature chaos) and memory effects.
Moreover, because spin glasses are influential in so many other
fields (such as  econophysics, biology or optimization in computer science), our
work shows that successful modeling  of complex systems is feasible in
finite dimensions.

\begin{acknowledgments}
We are grateful for helpful discussions with S. Swinnea
about sample characterization. This work was partially
supported by the U.S. Department of Energy, Office of
Basic Energy Sciences, Division of Materials Science and
Engineering, under Award No. DE-SC0013599, and
Contract No. DE-AC02-07CH11358; by the Ministerio
de Econom\'ia, Industria y Competitividad (MINECO,
Spain), Agencia Estatal de Investigaci\'on (AEI, Spain),
and Fondo Europeo de Desarrollo Regional (FEDER, EU)
through Grants No. FIS2016-76359-P, No. PID2019-
103939RB-I00, No. PGC2018-094684-B-C21, and
No. PGC2018-094684-B-C22; by the Junta de
Extremadura (Spain) and Fondo Europeo de Desarrollo
Regional (FEDER, EU) through Grants No. GRU18079
and No. IB15013. This project has also received funding
from the European Research Council (ERC) under the
European Union’s Horizon 2020 research and innovation
program (Grant No. 694925-LotglasSy). D. Y. was supported by the Chan Zuckerberg Biohub
and I. G. A. P. was
supported by the Ministerio de Ciencia, Innovaci\'on y
Universidades (MCIU, Spain) through FPU Grant
No. FPU18/02665. B. S. was supported by the
Comunidad de Madrid and the Complutense University
of Madrid (Spain) through the Atracci\'on de Talento
program (Ref. 2019-T1/TIC-12776).
\end{acknowledgments}
\appendix
\section{Comparison between different non-linear scaling laws}
In the main text, we introduced the non-linear scaling law~\eqref{eq:prediction}, 
that we argued represents a significant step forward in Ref.~\citep{janus:17b}. 

In order to give a better sense of the improvement achieved through Eq.~\eqref{eq:prediction}, we show here that the data do not collapse equally well if we use two different scaling laws, one from Ref. \citep{janus:17b} and the other a simple rational modification of Ref. \citep{janus:17b}.
Specifically, we reanalyze our data both through the scaling equation proposed in Ref.~\cite{janus:17b},
\begin{equation}
\label{eq:scaling_2017}
\ln \frac{t^{\text{eff}}_H}{t^{\text{eff}}_{H\to 0^+}}={\cal F} \left( \xi^{D-\frac{\theta(\tilde{x})}{2}} H^2 \right) \, ,
\end{equation}
or by postulating that the data can be rationalized through a single scaling term,
\begin{equation}
\label{eq:scaling_referreB}
\left[\ln \frac{t^{\text{eff}}_H}{t^{\text{eff}}_{H\to 0^+}}\right] \xi^{\theta(\tilde{x})/2} ={\cal F} \left( \xi^{D-\frac{\theta(\tilde{x})}{2}} H^2 \right) \, .
\end{equation}

\begin{figure}[b]
	\centering
	\includegraphics[width = \columnwidth]{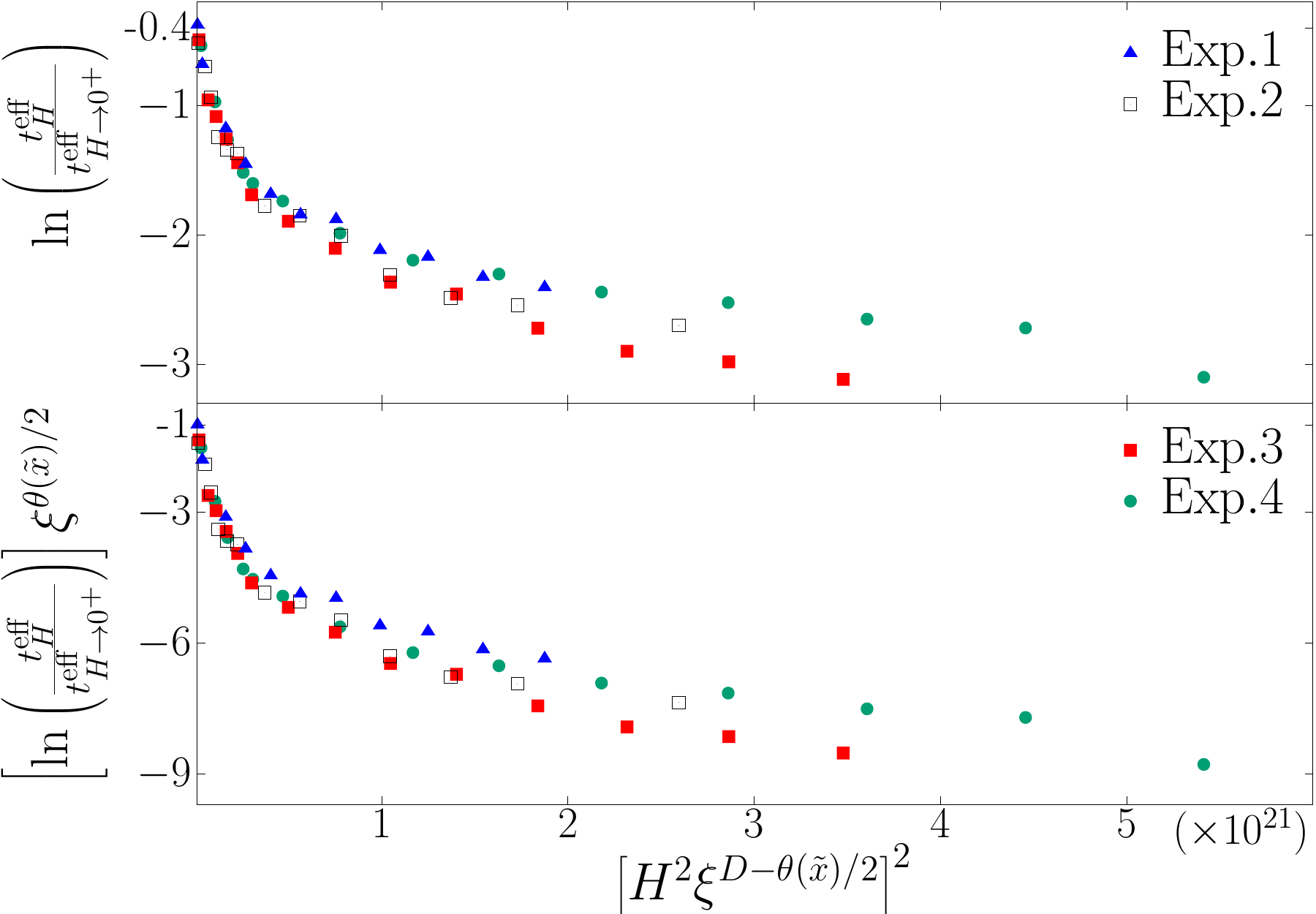}\vspace{-0.4cm}
	\caption{\label{fig:scaling_exp_2017_and_refeereB_Fig4} {\bf Top} The non-linear part of the experimental response time data:
	$[\ln (t_H^{\text{eff}}/t^{\text{eff}}_{H \to 0^+})$ plotted against the scaling variable           $[H^2\xi^{D-\theta(\tilde{x})/2}]^2$, according to Ref. \citep{janus:17b}, see Eq.~\eqref{eq:scaling_2017}. The plot is in linear scale.
          {\bf Bottom} The non-linear part of the response time data:$[\ln (
          t_H^{\text{eff}}/t^{\text{eff}}_{H \to 0^+}) \xi^{\theta(\tilde{x})/2}$ plotted against the scaling variable  $[H^2\xi^{D-\theta(\tilde{x})/2}]^2$, according to Eq.~\eqref{eq:scaling_referreB}. The plot is in linear scale.}
      \end{figure}

\begin{figure}[h]
	\centering
	\includegraphics[width = \columnwidth]{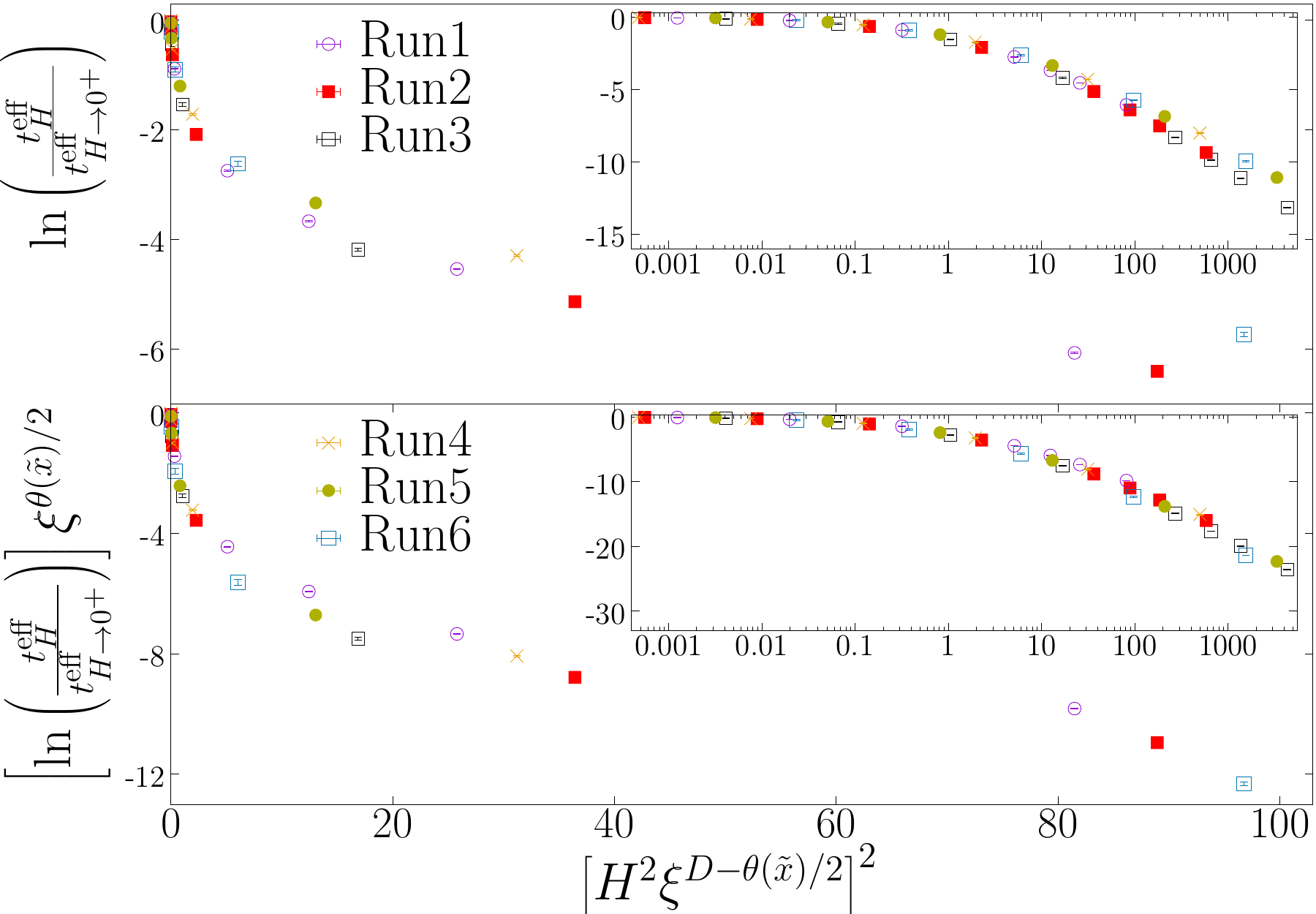}\vspace{-0.3cm}
	\caption{\label{fig:scaling_numerical_2017_and_refeereB_Fig4} {\bf Top} The non-linear part of the numerical response time data:
	$[\ln (t_H^{\text{eff}}/t^{\text{eff}}_{H \to 0^+})$ plotted against the scaling variable $[H^2\xi^{D-\theta(\tilde{x})/2}]^2$, according to Ref. \citep{janus:17b}, see Eq.~\eqref{eq:scaling_2017}. Its {\bf main} panel is in linear scale, instead its {\bf insert} is in semi-log scale.
          {\bf Bottom} The non-linear part of the response time data:$[\ln (
          t_H^{\text{eff}}/t^{\text{eff}}_{H \to 0^+}) \xi^{\theta(\tilde{x})/2}$ plotted against the scaling variable  $[H^2\xi^{D-\theta(\tilde{x})/2}]^2$, according to Eq.~\eqref{eq:scaling_referreB}. Its {\bf main} panel is in linear scale, instead its {\bf insert} is in semi-log scale. }
\end{figure}

\begin{figure}[h]
	\includegraphics[width = \columnwidth]{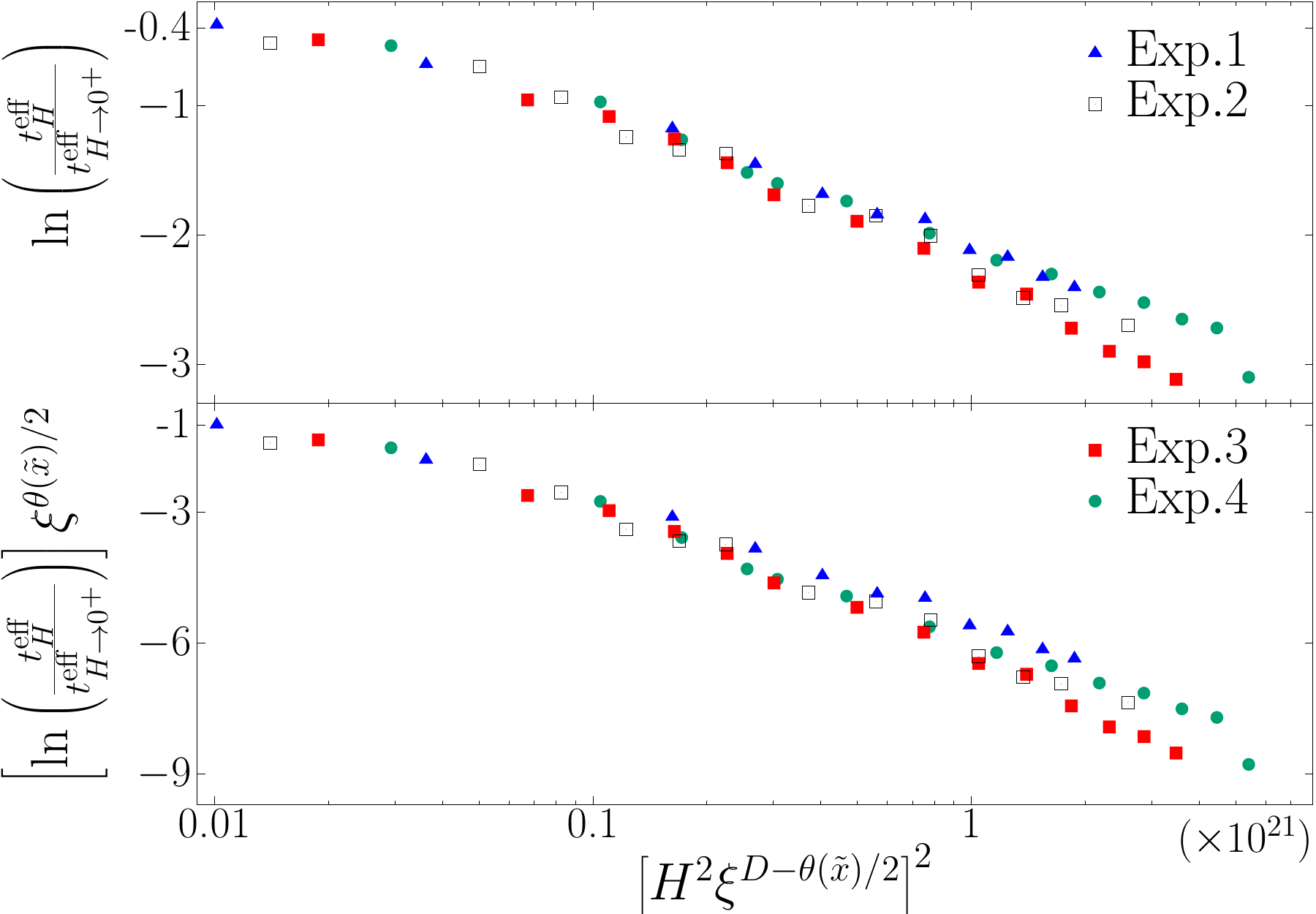}
	\caption{\label{fig:scaling_exp_logscale_2017_referreB}  {\bf Top} The non-linear part of the experimental response time data:
	$[\ln (t_H^{\text{eff}}/t^{\text{eff}}_{H \to 0^+})$ plotted against the scaling variable           $[H^2\xi^{D-\theta(\tilde{x})/2}]^2$, according to Ref. \citep{janus:17b}, see Eq.~\eqref{eq:scaling_2017}. The plot is in semi-log scale.
          {\bf Bottom} The non-linear part of the response time data: $[\ln (
          t_H^{\text{eff}}/t^{\text{eff}}_{H \to 0^+}) \xi^{\theta(\tilde{x})/2}$ plotted against the scaling variable  $[H^2\xi^{D-\theta(\tilde{x})/2}]^2$, according to Eq.~\eqref{eq:scaling_referreB}. The plot is in semi-log scale.}

	\centering
	\includegraphics[width = \columnwidth]{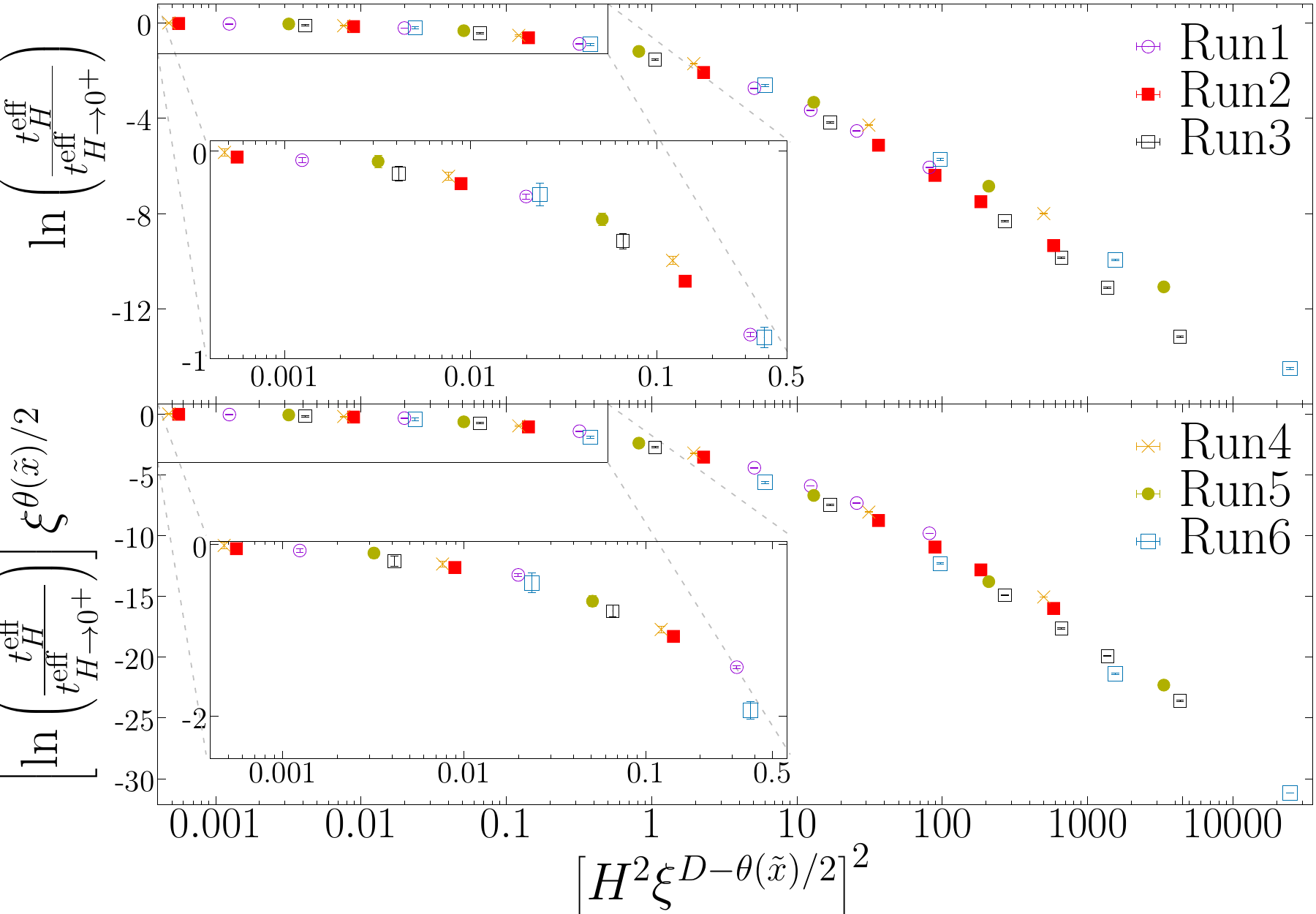}
	\caption{\label{fig:scaling_num_logscale_2017_and_referreB} {\bf Top} The non-linear part of the numerical response time data:
	$[\ln (t_H^{\text{eff}}/t^{\text{eff}}_{H \to 0^+})$ plotted against the scaling variable           $[H^2\xi^{D-\theta(\tilde{x})/2}]^2$, according to Ref. \citep{janus:17b}, see Eq.~\eqref{eq:scaling_2017}. Its {\bf main} panel is in semi-log scale and its {\bf insert} is a zoom in the small $x$ region.
          {\bf Bottom} The non-linear part of the response time data:$[\ln (
          t_H^{\text{eff}}/t^{\text{eff}}_{H \to 0^+}) \xi^{\theta(\tilde{x})/2}$ plotted against the scaling variable  $[H^2\xi^{D-\theta(\tilde{x})/2}]^2$, according to Eq.~\eqref{eq:scaling_referreB}. Its {\bf main} panel is in semi-log scale and its {\bf insert} is a zoom in the small x range. }
\end{figure}

We report the non-linear scaling behaviors using Eqs.~\eqref{eq:scaling_2017} and~\eqref{eq:scaling_referreB} in Figs.~\ref{fig:scaling_exp_2017_and_refeereB_Fig4} (experimental data) and~\ref{fig:scaling_numerical_2017_and_refeereB_Fig4} (numerical data).
To ease comparisons, we use the same scaling variable and $x$-axis scale that we used in Fig.~4 of the main text, where we collapsed the data using Eq.~\eqref{eq:prediction}.
However, because the scalings are easier to interpret with a linear $y$-axis scale, we also provide the same plots in a semi-log scale (Figs.~\ref{fig:scaling_exp_logscale_2017_referreB} and~\ref{fig:scaling_num_logscale_2017_and_referreB}).

The collapse of the experimental data with Eqs.~\eqref{eq:scaling_2017} and~\eqref{eq:scaling_referreB} (Figs.~\ref{fig:scaling_exp_2017_and_refeereB_Fig4} and \ref{fig:scaling_exp_logscale_2017_referreB}) 
works well only at most $x= (H^2 \xi^{D-\theta/2})^2= \, 6 \times 10^{20} \, \mathrm{Oe}$, which is about half of the validity range of the collapses in Fig.~4 of the main text, which are accurate at least up to $x=2 \times 10^{21} \, \mathrm{Oe}$.

The collapse of the numerical data with Eqs.~\eqref{eq:scaling_2017} and~\eqref{eq:scaling_referreB}  (Figs.~\ref{fig:scaling_numerical_2017_and_refeereB_Fig4} and \ref{fig:scaling_num_logscale_2017_and_referreB}) 
is less accurate throughout the whole range of $x$.\\
\indent We believe, therefore, that the scaling relationship represented by Eq. \ref{eq:prediction}, taken from the main text, is far superior.\\


\newpage
\begin{thebibliography}{41}%
\makeatletter
\providecommand \@ifxundefined [1]{%
 \@ifx{#1\undefined}
}%
\providecommand \@ifnum [1]{%
 \ifnum #1\expandafter \@firstoftwo
 \else \expandafter \@secondoftwo
 \fi
}%
\providecommand \@ifx [1]{%
 \ifx #1\expandafter \@firstoftwo
 \else \expandafter \@secondoftwo
 \fi
}%
\providecommand \natexlab [1]{#1}%
\providecommand \enquote  [1]{``#1''}%
\providecommand \bibnamefont  [1]{#1}%
\providecommand \bibfnamefont [1]{#1}%
\providecommand \citenamefont [1]{#1}%
\providecommand \href@noop [0]{\@secondoftwo}%
\providecommand \href [0]{\begingroup \@sanitize@url \@href}%
\providecommand \@href[1]{\@@startlink{#1}\@@href}%
\providecommand \@@href[1]{\endgroup#1\@@endlink}%
\providecommand \@sanitize@url [0]{\catcode `\\12\catcode `\$12\catcode
  `\&12\catcode `\#12\catcode `\^12\catcode `\_12\catcode `\%12\relax}%
\providecommand \@@startlink[1]{}%
\providecommand \@@endlink[0]{}%
\providecommand \url  [0]{\begingroup\@sanitize@url \@url }%
\providecommand \@url [1]{\endgroup\@href {#1}{\urlprefix }}%
\providecommand \urlprefix  [0]{URL }%
\providecommand \Eprint [0]{\href }%
\providecommand \doibase [0]{http://dx.doi.org/}%
\providecommand \selectlanguage [0]{\@gobble}%
\providecommand \bibinfo  [0]{\@secondoftwo}%
\providecommand \bibfield  [0]{\@secondoftwo}%
\providecommand \translation [1]{[#1]}%
\providecommand \BibitemOpen [0]{}%
\providecommand \bibitemStop [0]{}%
\providecommand \bibitemNoStop [0]{.\EOS\space}%
\providecommand \EOS [0]{\spacefactor3000\relax}%
\providecommand \BibitemShut  [1]{\csname bibitem#1\endcsname}%
\let\auto@bib@innerbib\@empty
\bibitem [{\citenamefont {Cavagna}(2009)}]{cavagna:09}%
  \BibitemOpen
  \bibfield  {author} {\bibinfo {author} {\bibfnamefont {A.}~\bibnamefont
  {Cavagna}},\ }\href {\doibase 10.1016/j.physrep.2009.03.003} {\bibfield
  {journal} {\bibinfo  {journal} {Physics Reports}\ }\textbf {\bibinfo {volume}
  {476}},\ \bibinfo {pages} {51} (\bibinfo {year} {2009})},\ \Eprint
  {http://arxiv.org/abs/arXiv:0903.4264} {arXiv:0903.4264} \BibitemShut
  {NoStop}%
\bibitem [{\citenamefont {Charbonneau}\ \emph {et~al.}(2014)\citenamefont
  {Charbonneau}, \citenamefont {Kurchan}, \citenamefont {Parisi}, \citenamefont
  {Urbani},\ and\ \citenamefont {Zamponi}}]{charbonneau:14}%
  \BibitemOpen
  \bibfield  {author} {\bibinfo {author} {\bibfnamefont {P.}~\bibnamefont
  {Charbonneau}}, \bibinfo {author} {\bibfnamefont {J.}~\bibnamefont
  {Kurchan}}, \bibinfo {author} {\bibfnamefont {G.}~\bibnamefont {Parisi}},
  \bibinfo {author} {\bibfnamefont {P.}~\bibnamefont {Urbani}}, \ and\ \bibinfo
  {author} {\bibfnamefont {F.}~\bibnamefont {Zamponi}},\ }\href {\doibase
  10.1038/ncomms4725} {\bibfield  {journal} {\bibinfo  {journal} {Nature
  Communications}\ }\textbf {\bibinfo {volume} {5}},\ \bibinfo {pages} {3725}
  (\bibinfo {year} {2014})},\ \Eprint {http://arxiv.org/abs/arXiv:1404.6809}
  {arXiv:1404.6809} \BibitemShut {NoStop}%
\bibitem [{\citenamefont {Adam}\ and\ \citenamefont {Gibbs}(1965)}]{adam:65}%
  \BibitemOpen
  \bibfield  {author} {\bibinfo {author} {\bibfnamefont {G.}~\bibnamefont
  {Adam}}\ and\ \bibinfo {author} {\bibfnamefont {J.~H.}\ \bibnamefont
  {Gibbs}},\ }\href {\doibase http://dx.doi.org/10.1063/1.1696442} {\bibfield
  {journal} {\bibinfo  {journal} {The Journal of Chemical Physics}\ }\textbf
  {\bibinfo {volume} {43}},\ \bibinfo {pages} {139} (\bibinfo {year}
  {1965})}\BibitemShut {NoStop}%
\bibitem [{\citenamefont {Marinari}\ \emph {et~al.}(1996)\citenamefont
  {Marinari}, \citenamefont {Parisi}, \citenamefont {Ruiz-Lorenzo},\ and\
  \citenamefont {Ritort}}]{marinari:96}%
  \BibitemOpen
  \bibfield  {author} {\bibinfo {author} {\bibfnamefont {E.}~\bibnamefont
  {Marinari}}, \bibinfo {author} {\bibfnamefont {G.}~\bibnamefont {Parisi}},
  \bibinfo {author} {\bibfnamefont {J.}~\bibnamefont {Ruiz-Lorenzo}}, \ and\
  \bibinfo {author} {\bibfnamefont {F.}~\bibnamefont {Ritort}},\ }\href
  {\doibase 10.1103/PhysRevLett.76.843} {\bibfield  {journal} {\bibinfo
  {journal} {Phys. Rev. Lett.}\ }\textbf {\bibinfo {volume} {76}},\ \bibinfo
  {pages} {843} (\bibinfo {year} {1996})}\BibitemShut {NoStop}%
\bibitem [{\citenamefont {Belletti}\ \emph {et~al.}(2008)\citenamefont
  {Belletti}, \citenamefont {Cotallo}, \citenamefont {Cruz}, \citenamefont
  {Fernandez}, \citenamefont {Gordillo-Guerrero}, \citenamefont {Guidetti},
  \citenamefont {Maiorano}, \citenamefont {Mantovani}, \citenamefont
  {Marinari}, \citenamefont {Mart\'{i}n-Mayor}, \citenamefont {Sudupe},
  \citenamefont {Navarro}, \citenamefont {Parisi}, \citenamefont
  {Perez-Gaviro}, \citenamefont {Ruiz-Lorenzo}, \citenamefont {Schifano},
  \citenamefont {Sciretti}, \citenamefont {Tarancon}, \citenamefont
  {Tripiccione}, \citenamefont {Velasco},\ and\ \citenamefont
  {Yllanes}}]{janus:08b}%
  \BibitemOpen
  \bibfield  {author} {\bibinfo {author} {\bibfnamefont {F.}~\bibnamefont
  {Belletti}}, \bibinfo {author} {\bibfnamefont {M.}~\bibnamefont {Cotallo}},
  \bibinfo {author} {\bibfnamefont {A.}~\bibnamefont {Cruz}}, \bibinfo {author}
  {\bibfnamefont {L.~A.}\ \bibnamefont {Fernandez}}, \bibinfo {author}
  {\bibfnamefont {A.}~\bibnamefont {Gordillo-Guerrero}}, \bibinfo {author}
  {\bibfnamefont {M.}~\bibnamefont {Guidetti}}, \bibinfo {author}
  {\bibfnamefont {A.}~\bibnamefont {Maiorano}}, \bibinfo {author}
  {\bibfnamefont {F.}~\bibnamefont {Mantovani}}, \bibinfo {author}
  {\bibfnamefont {E.}~\bibnamefont {Marinari}}, \bibinfo {author}
  {\bibfnamefont {V.}~\bibnamefont {Mart\'{i}n-Mayor}}, \bibinfo {author}
  {\bibfnamefont {A.~M.}\ \bibnamefont {Sudupe}}, \bibinfo {author}
  {\bibfnamefont {D.}~\bibnamefont {Navarro}}, \bibinfo {author} {\bibfnamefont
  {G.}~\bibnamefont {Parisi}}, \bibinfo {author} {\bibfnamefont
  {S.}~\bibnamefont {Perez-Gaviro}}, \bibinfo {author} {\bibfnamefont {J.~J.}\
  \bibnamefont {Ruiz-Lorenzo}}, \bibinfo {author} {\bibfnamefont {S.~F.}\
  \bibnamefont {Schifano}}, \bibinfo {author} {\bibfnamefont {D.}~\bibnamefont
  {Sciretti}}, \bibinfo {author} {\bibfnamefont {A.}~\bibnamefont {Tarancon}},
  \bibinfo {author} {\bibfnamefont {R.}~\bibnamefont {Tripiccione}}, \bibinfo
  {author} {\bibfnamefont {J.~L.}\ \bibnamefont {Velasco}}, \ and\ \bibinfo
  {author} {\bibfnamefont {D.}~\bibnamefont {Yllanes}} (\bibinfo
  {collaboration} {Janus Collaboration}),\ }\href {\doibase
  10.1103/PhysRevLett.101.157201} {\bibfield  {journal} {\bibinfo  {journal}
  {Phys. Rev. Lett.}\ }\textbf {\bibinfo {volume} {101}},\ \bibinfo {pages}
  {157201} (\bibinfo {year} {2008})},\ \Eprint
  {http://arxiv.org/abs/arXiv:0804.1471} {arXiv:0804.1471} \BibitemShut
  {NoStop}%
\bibitem [{\citenamefont {Belletti}\ \emph {et~al.}(2009)\citenamefont
  {Belletti}, \citenamefont {Cruz}, \citenamefont {Fernandez}, \citenamefont
  {Gordillo-Guerrero}, \citenamefont {Guidetti}, \citenamefont {Maiorano},
  \citenamefont {Mantovani}, \citenamefont {Marinari}, \citenamefont
  {Mart\'{i}n-Mayor}, \citenamefont {Monforte}, \citenamefont
  {Mu{\~n}oz~Sudupe}, \citenamefont {Navarro}, \citenamefont {Parisi},
  \citenamefont {Perez-Gaviro}, \citenamefont {Ruiz-Lorenzo}, \citenamefont
  {Schifano}, \citenamefont {Sciretti}, \citenamefont {Tarancon}, \citenamefont
  {Tripiccione},\ and\ \citenamefont {Yllanes}}]{janus:09b}%
  \BibitemOpen
  \bibfield  {author} {\bibinfo {author} {\bibfnamefont {F.}~\bibnamefont
  {Belletti}}, \bibinfo {author} {\bibfnamefont {A.}~\bibnamefont {Cruz}},
  \bibinfo {author} {\bibfnamefont {L.~A.}\ \bibnamefont {Fernandez}}, \bibinfo
  {author} {\bibfnamefont {A.}~\bibnamefont {Gordillo-Guerrero}}, \bibinfo
  {author} {\bibfnamefont {M.}~\bibnamefont {Guidetti}}, \bibinfo {author}
  {\bibfnamefont {A.}~\bibnamefont {Maiorano}}, \bibinfo {author}
  {\bibfnamefont {F.}~\bibnamefont {Mantovani}}, \bibinfo {author}
  {\bibfnamefont {E.}~\bibnamefont {Marinari}}, \bibinfo {author}
  {\bibfnamefont {V.}~\bibnamefont {Mart\'{i}n-Mayor}}, \bibinfo {author}
  {\bibfnamefont {J.}~\bibnamefont {Monforte}}, \bibinfo {author}
  {\bibfnamefont {A.}~\bibnamefont {Mu{\~n}oz~Sudupe}}, \bibinfo {author}
  {\bibfnamefont {D.}~\bibnamefont {Navarro}}, \bibinfo {author} {\bibfnamefont
  {G.}~\bibnamefont {Parisi}}, \bibinfo {author} {\bibfnamefont
  {S.}~\bibnamefont {Perez-Gaviro}}, \bibinfo {author} {\bibfnamefont {J.~J.}\
  \bibnamefont {Ruiz-Lorenzo}}, \bibinfo {author} {\bibfnamefont {S.~F.}\
  \bibnamefont {Schifano}}, \bibinfo {author} {\bibfnamefont {D.}~\bibnamefont
  {Sciretti}}, \bibinfo {author} {\bibfnamefont {A.}~\bibnamefont {Tarancon}},
  \bibinfo {author} {\bibfnamefont {R.}~\bibnamefont {Tripiccione}}, \ and\
  \bibinfo {author} {\bibfnamefont {D.}~\bibnamefont {Yllanes}} (\bibinfo
  {collaboration} {Janus Collaboration}),\ }\href {\doibase
  10.1007/s10955-009-9727-z} {\bibfield  {journal} {\bibinfo  {journal} {J.
  Stat. Phys.}\ }\textbf {\bibinfo {volume} {135}},\ \bibinfo {pages} {1121}
  (\bibinfo {year} {2009})},\ \Eprint {http://arxiv.org/abs/arXiv:0811.2864}
  {arXiv:0811.2864} \BibitemShut {NoStop}%
\bibitem [{\citenamefont {Alvarez~Ba{\~n}os}\ \emph {et~al.}(2010)\citenamefont
  {Alvarez~Ba{\~n}os}, \citenamefont {Cruz}, \citenamefont {Fernandez},
  \citenamefont {Gil-Narvion}, \citenamefont {Gordillo-Guerrero}, \citenamefont
  {Guidetti}, \citenamefont {Maiorano}, \citenamefont {Mantovani},
  \citenamefont {Marinari}, \citenamefont {Mart\'{i}n-Mayor}, \citenamefont
  {Monforte-Garcia}, \citenamefont {Mu{\~n}oz~Sudupe}, \citenamefont {Navarro},
  \citenamefont {Parisi}, \citenamefont {Perez-Gaviro}, \citenamefont
  {Ruiz-Lorenzo}, \citenamefont {Schifano}, \citenamefont {Seoane},
  \citenamefont {Tarancon}, \citenamefont {Tripiccione},\ and\ \citenamefont
  {Yllanes}}]{janus:10b}%
  \BibitemOpen
  \bibfield  {author} {\bibinfo {author} {\bibfnamefont {R.}~\bibnamefont
  {Alvarez~Ba{\~n}os}}, \bibinfo {author} {\bibfnamefont {A.}~\bibnamefont
  {Cruz}}, \bibinfo {author} {\bibfnamefont {L.~A.}\ \bibnamefont {Fernandez}},
  \bibinfo {author} {\bibfnamefont {J.~M.}\ \bibnamefont {Gil-Narvion}},
  \bibinfo {author} {\bibfnamefont {A.}~\bibnamefont {Gordillo-Guerrero}},
  \bibinfo {author} {\bibfnamefont {M.}~\bibnamefont {Guidetti}}, \bibinfo
  {author} {\bibfnamefont {A.}~\bibnamefont {Maiorano}}, \bibinfo {author}
  {\bibfnamefont {F.}~\bibnamefont {Mantovani}}, \bibinfo {author}
  {\bibfnamefont {E.}~\bibnamefont {Marinari}}, \bibinfo {author}
  {\bibfnamefont {V.}~\bibnamefont {Mart\'{i}n-Mayor}}, \bibinfo {author}
  {\bibfnamefont {J.}~\bibnamefont {Monforte-Garcia}}, \bibinfo {author}
  {\bibfnamefont {A.}~\bibnamefont {Mu{\~n}oz~Sudupe}}, \bibinfo {author}
  {\bibfnamefont {D.}~\bibnamefont {Navarro}}, \bibinfo {author} {\bibfnamefont
  {G.}~\bibnamefont {Parisi}}, \bibinfo {author} {\bibfnamefont
  {S.}~\bibnamefont {Perez-Gaviro}}, \bibinfo {author} {\bibfnamefont {J.~J.}\
  \bibnamefont {Ruiz-Lorenzo}}, \bibinfo {author} {\bibfnamefont {S.~F.}\
  \bibnamefont {Schifano}}, \bibinfo {author} {\bibfnamefont {B.}~\bibnamefont
  {Seoane}}, \bibinfo {author} {\bibfnamefont {A.}~\bibnamefont {Tarancon}},
  \bibinfo {author} {\bibfnamefont {R.}~\bibnamefont {Tripiccione}}, \ and\
  \bibinfo {author} {\bibfnamefont {D.}~\bibnamefont {Yllanes}} (\bibinfo
  {collaboration} {Janus Collaboration}),\ }\href {\doibase
  10.1103/PhysRevLett.105.177202} {\bibfield  {journal} {\bibinfo  {journal}
  {Phys. Rev. Lett.}\ }\textbf {\bibinfo {volume} {105}},\ \bibinfo {pages}
  {177202} (\bibinfo {year} {2010})},\ \Eprint
  {http://arxiv.org/abs/arXiv:1003.2943} {arXiv:1003.2943} \BibitemShut
  {NoStop}%
\bibitem [{\citenamefont {Manssen}\ and\ \citenamefont
  {Hartmann}(2015)}]{manssen:15}%
  \BibitemOpen
  \bibfield  {author} {\bibinfo {author} {\bibfnamefont {M.}~\bibnamefont
  {Manssen}}\ and\ \bibinfo {author} {\bibfnamefont {A.~K.}\ \bibnamefont
  {Hartmann}},\ }\href {\doibase 10.1103/PhysRevB.91.174433} {\bibfield
  {journal} {\bibinfo  {journal} {Phys. Rev. B}\ }\textbf {\bibinfo {volume}
  {91}},\ \bibinfo {pages} {174433} (\bibinfo {year} {2015})},\ \Eprint
  {http://arxiv.org/abs/arXiv:1411.5512} {arXiv:1411.5512} \BibitemShut
  {NoStop}%
\bibitem [{\citenamefont {Manssen}\ \emph {et~al.}(2015)\citenamefont
  {Manssen}, \citenamefont {Hartmann},\ and\ \citenamefont
  {Young}}]{manssen:15b}%
  \BibitemOpen
  \bibfield  {author} {\bibinfo {author} {\bibfnamefont {M.}~\bibnamefont
  {Manssen}}, \bibinfo {author} {\bibfnamefont {A.~K.}\ \bibnamefont
  {Hartmann}}, \ and\ \bibinfo {author} {\bibfnamefont {A.~P.}\ \bibnamefont
  {Young}},\ }\href {\doibase 10.1103/PhysRevB.91.104430} {\bibfield  {journal}
  {\bibinfo  {journal} {Phys. Rev. B}\ }\textbf {\bibinfo {volume} {91}},\
  \bibinfo {pages} {104430} (\bibinfo {year} {2015})},\ \Eprint
  {http://arxiv.org/abs/arXiv:1501.06760} {arXiv:1501.06760} \BibitemShut
  {NoStop}%
\bibitem [{\citenamefont {Fern\'andez}\ and\ \citenamefont
  {Mart\'{i}n-Mayor}(2015)}]{fernandez:15}%
  \BibitemOpen
  \bibfield  {author} {\bibinfo {author} {\bibfnamefont {L.~A.}\ \bibnamefont
  {Fern\'andez}}\ and\ \bibinfo {author} {\bibfnamefont {V.}~\bibnamefont
  {Mart\'{i}n-Mayor}},\ }\href {\doibase 10.1103/PhysRevB.91.174202} {\bibfield
   {journal} {\bibinfo  {journal} {Phys. Rev. B}\ }\textbf {\bibinfo {volume}
  {91}},\ \bibinfo {pages} {174202} (\bibinfo {year} {2015})}\BibitemShut
  {NoStop}%
\bibitem [{\citenamefont {Lulli}\ \emph {et~al.}(2016)\citenamefont {Lulli},
  \citenamefont {Parisi},\ and\ \citenamefont {Pelissetto}}]{lulli:15}%
  \BibitemOpen
  \bibfield  {author} {\bibinfo {author} {\bibfnamefont {M.}~\bibnamefont
  {Lulli}}, \bibinfo {author} {\bibfnamefont {G.}~\bibnamefont {Parisi}}, \
  and\ \bibinfo {author} {\bibfnamefont {A.}~\bibnamefont {Pelissetto}},\
  }\href {\doibase 10.1103/PhysRevE.93.032126} {\bibfield  {journal} {\bibinfo
  {journal} {Phys. Rev. E}\ }\textbf {\bibinfo {volume} {93}},\ \bibinfo
  {pages} {032126} (\bibinfo {year} {2016})}\BibitemShut {NoStop}%
\bibitem [{\citenamefont {Baity-Jesi}\ \emph
  {et~al.}(2017{\natexlab{a}})\citenamefont {Baity-Jesi}, \citenamefont
  {Calore}, \citenamefont {Cruz}, \citenamefont {Fernandez}, \citenamefont
  {Gil-Narvion}, \citenamefont {Gordillo-Guerrero}, \citenamefont {I\~niguez},
  \citenamefont {Maiorano}, \citenamefont {Marinari}, \citenamefont
  {Martin-Mayor}, \citenamefont {Monforte-Garcia}, \citenamefont {Mu\~noz
  Sudupe}, \citenamefont {Navarro}, \citenamefont {Parisi}, \citenamefont
  {Perez-Gaviro}, \citenamefont {Ricci-Tersenghi}, \citenamefont
  {Ruiz-Lorenzo}, \citenamefont {Schifano}, \citenamefont {Seoane},
  \citenamefont {Tarancon}, \citenamefont {Tripiccione},\ and\ \citenamefont
  {Yllanes}}]{janus:17b}%
  \BibitemOpen
  \bibfield  {author} {\bibinfo {author} {\bibfnamefont {M.}~\bibnamefont
  {Baity-Jesi}}, \bibinfo {author} {\bibfnamefont {E.}~\bibnamefont {Calore}},
  \bibinfo {author} {\bibfnamefont {A.}~\bibnamefont {Cruz}}, \bibinfo {author}
  {\bibfnamefont {L.~A.}\ \bibnamefont {Fernandez}}, \bibinfo {author}
  {\bibfnamefont {J.~M.}\ \bibnamefont {Gil-Narvion}}, \bibinfo {author}
  {\bibfnamefont {A.}~\bibnamefont {Gordillo-Guerrero}}, \bibinfo {author}
  {\bibfnamefont {D.}~\bibnamefont {I\~niguez}}, \bibinfo {author}
  {\bibfnamefont {A.}~\bibnamefont {Maiorano}}, \bibinfo {author}
  {\bibfnamefont {E.}~\bibnamefont {Marinari}}, \bibinfo {author}
  {\bibfnamefont {V.}~\bibnamefont {Martin-Mayor}}, \bibinfo {author}
  {\bibfnamefont {J.}~\bibnamefont {Monforte-Garcia}}, \bibinfo {author}
  {\bibfnamefont {A.}~\bibnamefont {Mu\~noz Sudupe}}, \bibinfo {author}
  {\bibfnamefont {D.}~\bibnamefont {Navarro}}, \bibinfo {author} {\bibfnamefont
  {G.}~\bibnamefont {Parisi}}, \bibinfo {author} {\bibfnamefont
  {S.}~\bibnamefont {Perez-Gaviro}}, \bibinfo {author} {\bibfnamefont
  {F.}~\bibnamefont {Ricci-Tersenghi}}, \bibinfo {author} {\bibfnamefont
  {J.~J.}\ \bibnamefont {Ruiz-Lorenzo}}, \bibinfo {author} {\bibfnamefont
  {S.~F.}\ \bibnamefont {Schifano}}, \bibinfo {author} {\bibfnamefont
  {B.}~\bibnamefont {Seoane}}, \bibinfo {author} {\bibfnamefont
  {A.}~\bibnamefont {Tarancon}}, \bibinfo {author} {\bibfnamefont
  {R.}~\bibnamefont {Tripiccione}}, \ and\ \bibinfo {author} {\bibfnamefont
  {D.}~\bibnamefont {Yllanes}} (\bibinfo {collaboration} {Janus
  Collaboration}),\ }\href {\doibase 10.1103/PhysRevLett.118.157202} {\bibfield
   {journal} {\bibinfo  {journal} {Phys. Rev. Lett.}\ }\textbf {\bibinfo
  {volume} {118}},\ \bibinfo {pages} {157202} (\bibinfo {year}
  {2017}{\natexlab{a}})}\BibitemShut {NoStop}%
\bibitem [{\citenamefont {Baity-Jesi}\ \emph {et~al.}(2018)\citenamefont
  {Baity-Jesi}, \citenamefont {Calore}, \citenamefont {Cruz}, \citenamefont
  {Fernandez}, \citenamefont {Gil-Narvion}, \citenamefont {Gordillo-Guerrero},
  \citenamefont {I\~niguez}, \citenamefont {Maiorano}, \citenamefont
  {Marinari}, \citenamefont {Martin-Mayor}, \citenamefont {Moreno-Gordo},
  \citenamefont {Mu\~noz Sudupe}, \citenamefont {Navarro}, \citenamefont
  {Parisi}, \citenamefont {Perez-Gaviro}, \citenamefont {Ricci-Tersenghi},
  \citenamefont {Ruiz-Lorenzo}, \citenamefont {Schifano}, \citenamefont
  {Seoane}, \citenamefont {Tarancon}, \citenamefont {Tripiccione},\ and\
  \citenamefont {Yllanes}}]{janus:18}%
  \BibitemOpen
  \bibfield  {author} {\bibinfo {author} {\bibfnamefont {M.}~\bibnamefont
  {Baity-Jesi}}, \bibinfo {author} {\bibfnamefont {E.}~\bibnamefont {Calore}},
  \bibinfo {author} {\bibfnamefont {A.}~\bibnamefont {Cruz}}, \bibinfo {author}
  {\bibfnamefont {L.~A.}\ \bibnamefont {Fernandez}}, \bibinfo {author}
  {\bibfnamefont {J.~M.}\ \bibnamefont {Gil-Narvion}}, \bibinfo {author}
  {\bibfnamefont {A.}~\bibnamefont {Gordillo-Guerrero}}, \bibinfo {author}
  {\bibfnamefont {D.}~\bibnamefont {I\~niguez}}, \bibinfo {author}
  {\bibfnamefont {A.}~\bibnamefont {Maiorano}}, \bibinfo {author}
  {\bibfnamefont {E.}~\bibnamefont {Marinari}}, \bibinfo {author}
  {\bibfnamefont {V.}~\bibnamefont {Martin-Mayor}}, \bibinfo {author}
  {\bibfnamefont {J.}~\bibnamefont {Moreno-Gordo}}, \bibinfo {author}
  {\bibfnamefont {A.}~\bibnamefont {Mu\~noz Sudupe}}, \bibinfo {author}
  {\bibfnamefont {D.}~\bibnamefont {Navarro}}, \bibinfo {author} {\bibfnamefont
  {G.}~\bibnamefont {Parisi}}, \bibinfo {author} {\bibfnamefont
  {S.}~\bibnamefont {Perez-Gaviro}}, \bibinfo {author} {\bibfnamefont
  {F.}~\bibnamefont {Ricci-Tersenghi}}, \bibinfo {author} {\bibfnamefont
  {J.~J.}\ \bibnamefont {Ruiz-Lorenzo}}, \bibinfo {author} {\bibfnamefont
  {S.~F.}\ \bibnamefont {Schifano}}, \bibinfo {author} {\bibfnamefont
  {B.}~\bibnamefont {Seoane}}, \bibinfo {author} {\bibfnamefont
  {A.}~\bibnamefont {Tarancon}}, \bibinfo {author} {\bibfnamefont
  {R.}~\bibnamefont {Tripiccione}}, \ and\ \bibinfo {author} {\bibfnamefont
  {D.}~\bibnamefont {Yllanes}} (\bibinfo {collaboration} {Janus
  Collaboration}),\ }\href {\doibase 10.1103/PhysRevLett.120.267203} {\bibfield
   {journal} {\bibinfo  {journal} {Phys. Rev. Lett.}\ }\textbf {\bibinfo
  {volume} {120}},\ \bibinfo {pages} {267203} (\bibinfo {year}
  {2018})}\BibitemShut {NoStop}%
\bibitem [{\citenamefont {Fernandez}\ \emph {et~al.}(2019)\citenamefont
  {Fernandez}, \citenamefont {Marinari}, \citenamefont {Martin-Mayor},
  \citenamefont {Paga},\ and\ \citenamefont {Ruiz-Lorenzo}}]{fernandez:19b}%
  \BibitemOpen
  \bibfield  {author} {\bibinfo {author} {\bibfnamefont {L.~A.}\ \bibnamefont
  {Fernandez}}, \bibinfo {author} {\bibfnamefont {E.}~\bibnamefont {Marinari}},
  \bibinfo {author} {\bibfnamefont {V.}~\bibnamefont {Martin-Mayor}}, \bibinfo
  {author} {\bibfnamefont {I.}~\bibnamefont {Paga}}, \ and\ \bibinfo {author}
  {\bibfnamefont {J.~J.}\ \bibnamefont {Ruiz-Lorenzo}},\ }\href {\doibase
  10.1103/PhysRevB.100.184412} {\bibfield  {journal} {\bibinfo  {journal}
  {Phys. Rev. B}\ }\textbf {\bibinfo {volume} {100}},\ \bibinfo {pages}
  {184412} (\bibinfo {year} {2019})}\BibitemShut {NoStop}%
\bibitem [{\citenamefont {Albert}\ \emph {et~al.}(2016)\citenamefont {Albert},
  \citenamefont {Bauer}, \citenamefont {Michl}, \citenamefont {Biroli},
  \citenamefont {Bouchaud}, \citenamefont {Loidl}, \citenamefont
  {Lunkenheimer}, \citenamefont {Tourbot}, \citenamefont {Wiertel-Gasquet},\
  and\ \citenamefont {Ladieu}}]{albert:16}%
  \BibitemOpen
  \bibfield  {author} {\bibinfo {author} {\bibfnamefont {S.}~\bibnamefont
  {Albert}}, \bibinfo {author} {\bibfnamefont {T.}~\bibnamefont {Bauer}},
  \bibinfo {author} {\bibfnamefont {M.}~\bibnamefont {Michl}}, \bibinfo
  {author} {\bibfnamefont {G.}~\bibnamefont {Biroli}}, \bibinfo {author}
  {\bibfnamefont {J.-P.}\ \bibnamefont {Bouchaud}}, \bibinfo {author}
  {\bibfnamefont {A.}~\bibnamefont {Loidl}}, \bibinfo {author} {\bibfnamefont
  {P.}~\bibnamefont {Lunkenheimer}}, \bibinfo {author} {\bibfnamefont
  {R.}~\bibnamefont {Tourbot}}, \bibinfo {author} {\bibfnamefont
  {C.}~\bibnamefont {Wiertel-Gasquet}}, \ and\ \bibinfo {author} {\bibfnamefont
  {F.}~\bibnamefont {Ladieu}},\ }\href {\doibase 10.1126/science.aaf3182}
  {\bibfield  {journal} {\bibinfo  {journal} {Science}\ }\textbf {\bibinfo
  {volume} {352}},\ \bibinfo {pages} {1308} (\bibinfo {year} {2016})},\ \Eprint
  {http://arxiv.org/abs/arXiv:1606.04079} {arXiv:1606.04079} \BibitemShut
  {NoStop}%
\bibitem [{\citenamefont {Joh}\ \emph {et~al.}(1999)\citenamefont {Joh},
  \citenamefont {Orbach}, \citenamefont {Wood}, \citenamefont {Hammann},\ and\
  \citenamefont {Vincent}}]{joh:99}%
  \BibitemOpen
  \bibfield  {author} {\bibinfo {author} {\bibfnamefont {Y.~G.}\ \bibnamefont
  {Joh}}, \bibinfo {author} {\bibfnamefont {R.}~\bibnamefont {Orbach}},
  \bibinfo {author} {\bibfnamefont {G.~G.}\ \bibnamefont {Wood}}, \bibinfo
  {author} {\bibfnamefont {J.}~\bibnamefont {Hammann}}, \ and\ \bibinfo
  {author} {\bibfnamefont {E.}~\bibnamefont {Vincent}},\ }\href {\doibase
  10.1103/PhysRevLett.82.438} {\bibfield  {journal} {\bibinfo  {journal} {Phys.
  Rev. Lett.}\ }\textbf {\bibinfo {volume} {82}},\ \bibinfo {pages} {438}
  (\bibinfo {year} {1999})}\BibitemShut {NoStop}%
\bibitem [{\citenamefont {Bert}\ \emph {et~al.}(2004)\citenamefont {Bert},
  \citenamefont {Dupuis}, \citenamefont {Vincent}, \citenamefont {Hammann},\
  and\ \citenamefont {Bouchaud}}]{bert:04}%
  \BibitemOpen
  \bibfield  {author} {\bibinfo {author} {\bibfnamefont {F.}~\bibnamefont
  {Bert}}, \bibinfo {author} {\bibfnamefont {V.}~\bibnamefont {Dupuis}},
  \bibinfo {author} {\bibfnamefont {E.}~\bibnamefont {Vincent}}, \bibinfo
  {author} {\bibfnamefont {J.}~\bibnamefont {Hammann}}, \ and\ \bibinfo
  {author} {\bibfnamefont {J.-P.}\ \bibnamefont {Bouchaud}},\ }\href {\doibase
  10.1103/PhysRevLett.92.167203} {\bibfield  {journal} {\bibinfo  {journal}
  {Phys. Rev. Lett.}\ }\textbf {\bibinfo {volume} {92}},\ \bibinfo {pages}
  {167203} (\bibinfo {year} {2004})}\BibitemShut {NoStop}%
\bibitem [{\citenamefont {Guchhait}\ and\ \citenamefont
  {Orbach}(2014)}]{guchhait:14}%
  \BibitemOpen
  \bibfield  {author} {\bibinfo {author} {\bibfnamefont {S.}~\bibnamefont
  {Guchhait}}\ and\ \bibinfo {author} {\bibfnamefont {R.}~\bibnamefont
  {Orbach}},\ }\href {\doibase 10.1103/PhysRevLett.112.126401} {\bibfield
  {journal} {\bibinfo  {journal} {Phys. Rev. Lett.}\ }\textbf {\bibinfo
  {volume} {112}},\ \bibinfo {pages} {126401} (\bibinfo {year}
  {2014})}\BibitemShut {NoStop}%
\bibitem [{\citenamefont {Guchhait}\ and\ \citenamefont
  {Orbach}(2017)}]{guchhait:17}%
  \BibitemOpen
  \bibfield  {author} {\bibinfo {author} {\bibfnamefont {S.}~\bibnamefont
  {Guchhait}}\ and\ \bibinfo {author} {\bibfnamefont {R.~L.}\ \bibnamefont
  {Orbach}},\ }\href {\doibase 10.1103/PhysRevLett.118.157203} {\bibfield
  {journal} {\bibinfo  {journal} {Phys. Rev. Lett.}\ }\textbf {\bibinfo
  {volume} {118}},\ \bibinfo {pages} {157203} (\bibinfo {year}
  {2017})}\BibitemShut {NoStop}%
\bibitem [{\citenamefont {Zhai}\ \emph {et~al.}(2019)\citenamefont {Zhai},
  \citenamefont {Martin-Mayor}, \citenamefont {Schlagel}, \citenamefont
  {Kenning},\ and\ \citenamefont {Orbach}}]{zhai:19}%
  \BibitemOpen
  \bibfield  {author} {\bibinfo {author} {\bibfnamefont {Q.}~\bibnamefont
  {Zhai}}, \bibinfo {author} {\bibfnamefont {V.}~\bibnamefont {Martin-Mayor}},
  \bibinfo {author} {\bibfnamefont {D.~L.}\ \bibnamefont {Schlagel}}, \bibinfo
  {author} {\bibfnamefont {G.~G.}\ \bibnamefont {Kenning}}, \ and\ \bibinfo
  {author} {\bibfnamefont {R.~L.}\ \bibnamefont {Orbach}},\ }\href {\doibase
  10.1103/PhysRevB.100.094202} {\bibfield  {journal} {\bibinfo  {journal}
  {Phys. Rev. B}\ }\textbf {\bibinfo {volume} {100}},\ \bibinfo {pages}
  {094202} (\bibinfo {year} {2019})}\BibitemShut {NoStop}%
\bibitem [{\citenamefont {Baity-Jesi}\ \emph {et~al.}(2014)\citenamefont
  {Baity-Jesi}, \citenamefont {Ba\~{n}os}, \citenamefont {Cruz}, \citenamefont
  {Fernandez}, \citenamefont {Gil-Narvion}, \citenamefont {Gordillo-Guerrero},
  \citenamefont {Iniguez}, \citenamefont {Maiorano}, \citenamefont {Mantovani},
  \citenamefont {Marinari}, \citenamefont {Mart\'{i}n-Mayor}, \citenamefont
  {Monforte-Garcia}, \citenamefont {Mu{\~n}oz~Sudupe}, \citenamefont {Navarro},
  \citenamefont {Parisi}, \citenamefont {Perez-Gaviro}, \citenamefont
  {Pivanti}, \citenamefont {Ricci-Tersenghi}, \citenamefont {Ruiz-Lorenzo},
  \citenamefont {Schifano}, \citenamefont {Seoane}, \citenamefont {Tarancon},
  \citenamefont {Tripiccione},\ and\ \citenamefont {Yllanes}}]{janus:14}%
  \BibitemOpen
  \bibfield  {author} {\bibinfo {author} {\bibfnamefont {M.}~\bibnamefont
  {Baity-Jesi}}, \bibinfo {author} {\bibfnamefont {R.~A.}\ \bibnamefont
  {Ba\~{n}os}}, \bibinfo {author} {\bibfnamefont {A.}~\bibnamefont {Cruz}},
  \bibinfo {author} {\bibfnamefont {L.~A.}\ \bibnamefont {Fernandez}}, \bibinfo
  {author} {\bibfnamefont {J.~M.}\ \bibnamefont {Gil-Narvion}}, \bibinfo
  {author} {\bibfnamefont {A.}~\bibnamefont {Gordillo-Guerrero}}, \bibinfo
  {author} {\bibfnamefont {D.}~\bibnamefont {Iniguez}}, \bibinfo {author}
  {\bibfnamefont {A.}~\bibnamefont {Maiorano}}, \bibinfo {author}
  {\bibfnamefont {F.}~\bibnamefont {Mantovani}}, \bibinfo {author}
  {\bibfnamefont {E.}~\bibnamefont {Marinari}}, \bibinfo {author}
  {\bibfnamefont {V.}~\bibnamefont {Mart\'{i}n-Mayor}}, \bibinfo {author}
  {\bibfnamefont {J.}~\bibnamefont {Monforte-Garcia}}, \bibinfo {author}
  {\bibfnamefont {A.}~\bibnamefont {Mu{\~n}oz~Sudupe}}, \bibinfo {author}
  {\bibfnamefont {D.}~\bibnamefont {Navarro}}, \bibinfo {author} {\bibfnamefont
  {G.}~\bibnamefont {Parisi}}, \bibinfo {author} {\bibfnamefont
  {S.}~\bibnamefont {Perez-Gaviro}}, \bibinfo {author} {\bibfnamefont
  {M.}~\bibnamefont {Pivanti}}, \bibinfo {author} {\bibfnamefont
  {F.}~\bibnamefont {Ricci-Tersenghi}}, \bibinfo {author} {\bibfnamefont
  {J.~J.}\ \bibnamefont {Ruiz-Lorenzo}}, \bibinfo {author} {\bibfnamefont
  {S.~F.}\ \bibnamefont {Schifano}}, \bibinfo {author} {\bibfnamefont
  {B.}~\bibnamefont {Seoane}}, \bibinfo {author} {\bibfnamefont
  {A.}~\bibnamefont {Tarancon}}, \bibinfo {author} {\bibfnamefont
  {R.}~\bibnamefont {Tripiccione}}, \ and\ \bibinfo {author} {\bibfnamefont
  {D.}~\bibnamefont {Yllanes}} (\bibinfo {collaboration} {Janus
  Collaboration}),\ }\href {\doibase 10.1016/j.cpc.2013.10.019} {\bibfield
  {journal} {\bibinfo  {journal} {Comp. Phys. Comm}\ }\textbf {\bibinfo
  {volume} {185}},\ \bibinfo {pages} {550} (\bibinfo {year} {2014})},\ \Eprint
  {http://arxiv.org/abs/arXiv:1310.1032} {arXiv:1310.1032} \BibitemShut
  {NoStop}%
\bibitem [{\citenamefont {Vincent}\ \emph {et~al.}(1995)\citenamefont
  {Vincent}, \citenamefont {Bouchaud}, \citenamefont {Dean},\ and\
  \citenamefont {Hammann}}]{vincent:95}%
  \BibitemOpen
  \bibfield  {author} {\bibinfo {author} {\bibfnamefont {E.}~\bibnamefont
  {Vincent}}, \bibinfo {author} {\bibfnamefont {J.~P.}\ \bibnamefont
  {Bouchaud}}, \bibinfo {author} {\bibfnamefont {D.~S.}\ \bibnamefont {Dean}},
  \ and\ \bibinfo {author} {\bibfnamefont {J.}~\bibnamefont {Hammann}},\ }\href
  {\doibase 10.1103/PhysRevB.52.1050} {\bibfield  {journal} {\bibinfo
  {journal} {Phys. Rev. B}\ }\textbf {\bibinfo {volume} {52}},\ \bibinfo
  {pages} {1050} (\bibinfo {year} {1995})}\BibitemShut {NoStop}%
\bibitem [{\citenamefont {Kenning}\ \emph {et~al.}(2018)\citenamefont
  {Kenning}, \citenamefont {Tennant}, \citenamefont {Rost}, \citenamefont
  {da~Silva}, \citenamefont {Walters}, \citenamefont {Zhai}, \citenamefont
  {Harrison}, \citenamefont {Dahlberg},\ and\ \citenamefont
  {Orbach}}]{kenning:18}%
  \BibitemOpen
  \bibfield  {author} {\bibinfo {author} {\bibfnamefont {G.~G.}\ \bibnamefont
  {Kenning}}, \bibinfo {author} {\bibfnamefont {D.~M.}\ \bibnamefont
  {Tennant}}, \bibinfo {author} {\bibfnamefont {C.~M.}\ \bibnamefont {Rost}},
  \bibinfo {author} {\bibfnamefont {F.~G.}\ \bibnamefont {da~Silva}}, \bibinfo
  {author} {\bibfnamefont {B.~J.}\ \bibnamefont {Walters}}, \bibinfo {author}
  {\bibfnamefont {Q.}~\bibnamefont {Zhai}}, \bibinfo {author} {\bibfnamefont
  {D.~C.}\ \bibnamefont {Harrison}}, \bibinfo {author} {\bibfnamefont {E.~D.}\
  \bibnamefont {Dahlberg}}, \ and\ \bibinfo {author} {\bibfnamefont {R.~L.}\
  \bibnamefont {Orbach}},\ }\href {\doibase 10.1103/PhysRevB.98.104436}
  {\bibfield  {journal} {\bibinfo  {journal} {Phys. Rev. B}\ }\textbf {\bibinfo
  {volume} {98}},\ \bibinfo {pages} {104436} (\bibinfo {year}
  {2018})}\BibitemShut {NoStop}%
\bibitem [{Note1()}]{Note1}%
  \BibitemOpen
  \bibinfo {note} {In a $L=160$ system, $\xi (\protect \ensuremath {t_\protect
  \mathrm {w}}\protect \xspace )$ and $M(t,\protect \ensuremath {t_\protect
  \mathrm {w}}\protect \xspace ;H)$ display little sample dependence, see~\cite
  {zhai-janus:20b}. We have, however, run 512 independent thermal histories for
  our sample (the benefits of simulating many independent thermal histories are
  discussed in Ref.~\cite {janus:18})}\BibitemShut {NoStop}%
\bibitem [{\citenamefont {Baity-Jesi}\ \emph {et~al.}(2013)\citenamefont
  {Baity-Jesi}, \citenamefont {Ba\~{n}os}, \citenamefont {Cruz}, \citenamefont
  {Fernandez}, \citenamefont {Gil-Narvion}, \citenamefont {Gordillo-Guerrero},
  \citenamefont {Iniguez}, \citenamefont {Maiorano}, \citenamefont {Mantovani},
  \citenamefont {Marinari}, \citenamefont {Mart\'{i}n-Mayor}, \citenamefont
  {Monforte-Garcia}, \citenamefont {Mu{\~n}oz~Sudupe}, \citenamefont {Navarro},
  \citenamefont {Parisi}, \citenamefont {Perez-Gaviro}, \citenamefont
  {Pivanti}, \citenamefont {Ricci-Tersenghi}, \citenamefont {Ruiz-Lorenzo},
  \citenamefont {Schifano}, \citenamefont {Seoane}, \citenamefont {Tarancon},
  \citenamefont {Tripiccione},\ and\ \citenamefont {Yllanes}}]{janus:13}%
  \BibitemOpen
  \bibfield  {author} {\bibinfo {author} {\bibfnamefont {M.}~\bibnamefont
  {Baity-Jesi}}, \bibinfo {author} {\bibfnamefont {R.~A.}\ \bibnamefont
  {Ba\~{n}os}}, \bibinfo {author} {\bibfnamefont {A.}~\bibnamefont {Cruz}},
  \bibinfo {author} {\bibfnamefont {L.~A.}\ \bibnamefont {Fernandez}}, \bibinfo
  {author} {\bibfnamefont {J.~M.}\ \bibnamefont {Gil-Narvion}}, \bibinfo
  {author} {\bibfnamefont {A.}~\bibnamefont {Gordillo-Guerrero}}, \bibinfo
  {author} {\bibfnamefont {D.}~\bibnamefont {Iniguez}}, \bibinfo {author}
  {\bibfnamefont {A.}~\bibnamefont {Maiorano}}, \bibinfo {author}
  {\bibfnamefont {F.}~\bibnamefont {Mantovani}}, \bibinfo {author}
  {\bibfnamefont {E.}~\bibnamefont {Marinari}}, \bibinfo {author}
  {\bibfnamefont {V.}~\bibnamefont {Mart\'{i}n-Mayor}}, \bibinfo {author}
  {\bibfnamefont {J.}~\bibnamefont {Monforte-Garcia}}, \bibinfo {author}
  {\bibfnamefont {A.}~\bibnamefont {Mu{\~n}oz~Sudupe}}, \bibinfo {author}
  {\bibfnamefont {D.}~\bibnamefont {Navarro}}, \bibinfo {author} {\bibfnamefont
  {G.}~\bibnamefont {Parisi}}, \bibinfo {author} {\bibfnamefont
  {S.}~\bibnamefont {Perez-Gaviro}}, \bibinfo {author} {\bibfnamefont
  {M.}~\bibnamefont {Pivanti}}, \bibinfo {author} {\bibfnamefont
  {F.}~\bibnamefont {Ricci-Tersenghi}}, \bibinfo {author} {\bibfnamefont
  {J.~J.}\ \bibnamefont {Ruiz-Lorenzo}}, \bibinfo {author} {\bibfnamefont
  {S.~F.}\ \bibnamefont {Schifano}}, \bibinfo {author} {\bibfnamefont
  {B.}~\bibnamefont {Seoane}}, \bibinfo {author} {\bibfnamefont
  {A.}~\bibnamefont {Tarancon}}, \bibinfo {author} {\bibfnamefont
  {R.}~\bibnamefont {Tripiccione}}, \ and\ \bibinfo {author} {\bibfnamefont
  {D.}~\bibnamefont {Yllanes}} (\bibinfo {collaboration} {Janus
  Collaboration}),\ }\href {\doibase 10.1103/PhysRevB.88.224416} {\bibfield
  {journal} {\bibinfo  {journal} {Phys. Rev. B}\ }\textbf {\bibinfo {volume}
  {88}},\ \bibinfo {pages} {224416} (\bibinfo {year} {{2013}})},\ \Eprint
  {http://arxiv.org/abs/arXiv:1310.2910} {arXiv:1310.2910} \BibitemShut
  {NoStop}%
\bibitem [{\citenamefont {Aruga~Katori}\ and\ \citenamefont
  {Ito}(1994)}]{aruga_katori:94}%
  \BibitemOpen
  \bibfield  {author} {\bibinfo {author} {\bibfnamefont {H.}~\bibnamefont
  {Aruga~Katori}}\ and\ \bibinfo {author} {\bibfnamefont {A.}~\bibnamefont
  {Ito}},\ }\href {\doibase 10.1143/JPSJ.63.3122} {\bibfield  {journal}
  {\bibinfo  {journal} {Journal of the Physical Society of Japan}\ }\textbf
  {\bibinfo {volume} {63}},\ \bibinfo {pages} {3122} (\bibinfo {year}
  {1994})},\ \Eprint
  {http://arxiv.org/abs/http://dx.doi.org/10.1143/JPSJ.63.3122}
  {http://dx.doi.org/10.1143/JPSJ.63.3122} \BibitemShut {NoStop}%
\bibitem [{\citenamefont {Fisher}\ and\ \citenamefont
  {Sompolinsky}(1985)}]{fisher:85}%
  \BibitemOpen
  \bibfield  {author} {\bibinfo {author} {\bibfnamefont {D.~S.}\ \bibnamefont
  {Fisher}}\ and\ \bibinfo {author} {\bibfnamefont {H.}~\bibnamefont
  {Sompolinsky}},\ }\href {\doibase 10.1103/PhysRevLett.54.1063} {\bibfield
  {journal} {\bibinfo  {journal} {Phys. Rev. Lett.}\ }\textbf {\bibinfo
  {volume} {54}},\ \bibinfo {pages} {1063} (\bibinfo {year}
  {1985})}\BibitemShut {NoStop}%
\bibitem [{\citenamefont {Zhai}\ \emph {et~al.}(2020)\citenamefont {Zhai},
  \citenamefont {Schlagel}, \citenamefont {Orbach}, \citenamefont {Paga},
  \citenamefont {Baity-Jesi}, \citenamefont {Calore}, \citenamefont {Cruz},
  \citenamefont {Fernandez}, \citenamefont {Gil-Narvion}, \citenamefont
  {Gonz{\'{a}}lez-Adalid~Pemart{\'{\i}}n}, \citenamefont {Gordillo-Guerrero},
  \citenamefont {I\~niguez}, \citenamefont {Maiorano}, \citenamefont
  {Marinari}, \citenamefont {Martin-Mayor}, \citenamefont {Moreno-Gordo},
  \citenamefont {Mu\~noz Sudupe}, \citenamefont {Navarro}, \citenamefont
  {Parisi}, \citenamefont {Perez-Gaviro}, \citenamefont {Ricci-Tersenghi},
  \citenamefont {Ruiz-Lorenzo}, \citenamefont {Schifano}, \citenamefont
  {Seoane}, \citenamefont {Tarancon}, \citenamefont {Tripiccione},\ and\
  \citenamefont {Yllanes}}]{zhai-janus:20b}%
  \BibitemOpen
  \bibfield  {author} {\bibinfo {author} {\bibfnamefont {Q.}~\bibnamefont
  {Zhai}}, \bibinfo {author} {\bibfnamefont {D.~L.}\ \bibnamefont {Schlagel}},
  \bibinfo {author} {\bibfnamefont {R.~L.}\ \bibnamefont {Orbach}}, \bibinfo
  {author} {\bibfnamefont {I.}~\bibnamefont {Paga}}, \bibinfo {author}
  {\bibfnamefont {M.}~\bibnamefont {Baity-Jesi}}, \bibinfo {author}
  {\bibfnamefont {E.}~\bibnamefont {Calore}}, \bibinfo {author} {\bibfnamefont
  {A.}~\bibnamefont {Cruz}}, \bibinfo {author} {\bibfnamefont {L.~A.}\
  \bibnamefont {Fernandez}}, \bibinfo {author} {\bibfnamefont {J.~M.}\
  \bibnamefont {Gil-Narvion}}, \bibinfo {author} {\bibfnamefont
  {I.}~\bibnamefont {Gonz{\'{a}}lez-Adalid~Pemart{\'{\i}}n}}, \bibinfo {author}
  {\bibfnamefont {A.}~\bibnamefont {Gordillo-Guerrero}}, \bibinfo {author}
  {\bibfnamefont {D.}~\bibnamefont {I\~niguez}}, \bibinfo {author}
  {\bibfnamefont {A.}~\bibnamefont {Maiorano}}, \bibinfo {author}
  {\bibfnamefont {E.}~\bibnamefont {Marinari}}, \bibinfo {author}
  {\bibfnamefont {V.}~\bibnamefont {Martin-Mayor}}, \bibinfo {author}
  {\bibfnamefont {J.}~\bibnamefont {Moreno-Gordo}}, \bibinfo {author}
  {\bibfnamefont {A.}~\bibnamefont {Mu\~noz Sudupe}}, \bibinfo {author}
  {\bibfnamefont {D.}~\bibnamefont {Navarro}}, \bibinfo {author} {\bibfnamefont
  {G.}~\bibnamefont {Parisi}}, \bibinfo {author} {\bibfnamefont
  {S.}~\bibnamefont {Perez-Gaviro}}, \bibinfo {author} {\bibfnamefont
  {F.}~\bibnamefont {Ricci-Tersenghi}}, \bibinfo {author} {\bibfnamefont
  {J.~J.}\ \bibnamefont {Ruiz-Lorenzo}}, \bibinfo {author} {\bibfnamefont
  {S.~F.}\ \bibnamefont {Schifano}}, \bibinfo {author} {\bibfnamefont
  {B.}~\bibnamefont {Seoane}}, \bibinfo {author} {\bibfnamefont
  {A.}~\bibnamefont {Tarancon}}, \bibinfo {author} {\bibfnamefont
  {R.}~\bibnamefont {Tripiccione}}, \ and\ \bibinfo {author} {\bibfnamefont
  {D.}~\bibnamefont {Yllanes}},\ }\href@noop {} {} (\bibinfo {year} {2020}),\
  \bibinfo {note} {manuscript in preparation}\BibitemShut {NoStop}%
\bibitem [{\citenamefont {Gonz{\'{a}}lez-Adalid~Pemart{\'{\i}}n}\ \emph
  {et~al.}(2019)\citenamefont {Gonz{\'{a}}lez-Adalid~Pemart{\'{\i}}n},
  \citenamefont {Martin-Mayor}, \citenamefont {Parisi},\ and\ \citenamefont
  {Ruiz-Lorenzo}}]{gonzalez-adalid-pemartin:19}%
  \BibitemOpen
  \bibfield  {author} {\bibinfo {author} {\bibfnamefont {I.}~\bibnamefont
  {Gonz{\'{a}}lez-Adalid~Pemart{\'{\i}}n}}, \bibinfo {author} {\bibfnamefont
  {V.}~\bibnamefont {Martin-Mayor}}, \bibinfo {author} {\bibfnamefont
  {G.}~\bibnamefont {Parisi}}, \ and\ \bibinfo {author} {\bibfnamefont {J.~J.}\
  \bibnamefont {Ruiz-Lorenzo}},\ }\href {\doibase 10.1088/1751-8121/ab08d9}
  {\bibfield  {journal} {\bibinfo  {journal} {Journal of Physics A:
  Mathematical and Theoretical}\ }\textbf {\bibinfo {volume} {52}},\ \bibinfo
  {pages} {134002} (\bibinfo {year} {2019})}\BibitemShut {NoStop}%
\bibitem [{\citenamefont {Parisi}(1988)}]{parisi:88}%
  \BibitemOpen
  \bibfield  {author} {\bibinfo {author} {\bibfnamefont {G.}~\bibnamefont
  {Parisi}},\ }\href@noop {} {\emph {\bibinfo {title} {Statistical Field
  Theory}}}\ (\bibinfo  {publisher} {Addison-Wesley},\ \bibinfo {year}
  {1988})\BibitemShut {NoStop}%
\bibitem [{\citenamefont {Amit}\ and\ \citenamefont
  {Mart\'{i}n-Mayor}(2005)}]{amit:05}%
  \BibitemOpen
  \bibfield  {author} {\bibinfo {author} {\bibfnamefont {D.~J.}\ \bibnamefont
  {Amit}}\ and\ \bibinfo {author} {\bibfnamefont {V.}~\bibnamefont
  {Mart\'{i}n-Mayor}},\ }\href {\doibase 10.1142/9789812775313_bmatter} {\emph
  {\bibinfo {title} {Field Theory, the Renormalization Group and Critical
  Phenomena}}},\ \bibinfo {edition} {3rd}\ ed.\ (\bibinfo  {publisher} {World
  Scientific},\ \bibinfo {address} {Singapore},\ \bibinfo {year}
  {2005})\BibitemShut {NoStop}%
\bibitem [{\citenamefont {Rodriguez}\ \emph {et~al.}(2003)\citenamefont
  {Rodriguez}, \citenamefont {Kenning},\ and\ \citenamefont
  {Orbach}}]{rodriguez:03}%
  \BibitemOpen
  \bibfield  {author} {\bibinfo {author} {\bibfnamefont {G.~F.}\ \bibnamefont
  {Rodriguez}}, \bibinfo {author} {\bibfnamefont {G.~G.}\ \bibnamefont
  {Kenning}}, \ and\ \bibinfo {author} {\bibfnamefont {R.}~\bibnamefont
  {Orbach}},\ }\href {\doibase 10.1103/PhysRevLett.91.037203} {\bibfield
  {journal} {\bibinfo  {journal} {Phys. Rev. Lett.}\ }\textbf {\bibinfo
  {volume} {91}},\ \bibinfo {pages} {037203} (\bibinfo {year}
  {2003})}\BibitemShut {NoStop}%
\bibitem [{Note2()}]{Note2}%
  \BibitemOpen
  \bibinfo {note} {In order to not overburden the notation we have omitted the
  $t$ and $\protect \ensuremath {t_\protect \mathrm {w}}\protect \xspace $
  arguments in the r.h.s. susceptibilities in Eq.~\protect \textup {\hbox
  {\mathsurround \z@ \protect \normalfont (\ignorespaces \ref
  {eq:suscept-defined}\unskip \@@italiccorr )}}.}\BibitemShut {Stop}%
\bibitem [{\citenamefont {Cugliandolo}\ and\ \citenamefont
  {Kurchan}(1993)}]{cugliandolo:93}%
  \BibitemOpen
  \bibfield  {author} {\bibinfo {author} {\bibfnamefont {L.~F.}\ \bibnamefont
  {Cugliandolo}}\ and\ \bibinfo {author} {\bibfnamefont {J.}~\bibnamefont
  {Kurchan}},\ }\href {\doibase 10.1103/PhysRevLett.71.173} {\bibfield
  {journal} {\bibinfo  {journal} {Phys. Rev. Lett.}\ }\textbf {\bibinfo
  {volume} {71}},\ \bibinfo {pages} {173} (\bibinfo {year} {1993})}\BibitemShut
  {NoStop}%
\bibitem [{\citenamefont {Marinari}\ \emph {et~al.}(1998)\citenamefont
  {Marinari}, \citenamefont {Parisi}, \citenamefont {Ricci-Tersenghi},\ and\
  \citenamefont {Ruiz-Lorenzo}}]{marinari:98f}%
  \BibitemOpen
  \bibfield  {author} {\bibinfo {author} {\bibfnamefont {E.}~\bibnamefont
  {Marinari}}, \bibinfo {author} {\bibfnamefont {G.}~\bibnamefont {Parisi}},
  \bibinfo {author} {\bibfnamefont {F.}~\bibnamefont {Ricci-Tersenghi}}, \ and\
  \bibinfo {author} {\bibfnamefont {J.~J.}\ \bibnamefont {Ruiz-Lorenzo}},\
  }\href {\doibase 10.1088/0305-4470/31/11/011} {\bibfield  {journal} {\bibinfo
   {journal} {Journal of Physics A: Mathematical and General}\ }\textbf
  {\bibinfo {volume} {31}},\ \bibinfo {pages} {2611} (\bibinfo {year}
  {1998})}\BibitemShut {NoStop}%
\bibitem [{\citenamefont {Franz}\ \emph {et~al.}(1998)\citenamefont {Franz},
  \citenamefont {M\'ezard}, \citenamefont {Parisi},\ and\ \citenamefont
  {Peliti}}]{franz:98}%
  \BibitemOpen
  \bibfield  {author} {\bibinfo {author} {\bibfnamefont {S.}~\bibnamefont
  {Franz}}, \bibinfo {author} {\bibfnamefont {M.}~\bibnamefont {M\'ezard}},
  \bibinfo {author} {\bibfnamefont {G.}~\bibnamefont {Parisi}}, \ and\ \bibinfo
  {author} {\bibfnamefont {L.}~\bibnamefont {Peliti}},\ }\href {\doibase
  10.1103/PhysRevLett.81.1758} {\bibfield  {journal} {\bibinfo  {journal}
  {Phys. Rev. Lett.}\ }\textbf {\bibinfo {volume} {81}},\ \bibinfo {pages}
  {1758} (\bibinfo {year} {1998})}\BibitemShut {NoStop}%
\bibitem [{\citenamefont {Baity-Jesi}\ \emph
  {et~al.}(2017{\natexlab{b}})\citenamefont {Baity-Jesi}, \citenamefont
  {Calore}, \citenamefont {Cruz}, \citenamefont {Fernandez}, \citenamefont
  {Gil-Narvi\'on}, \citenamefont {Gordillo-Guerrero}, \citenamefont {Iñiguez},
  \citenamefont {Maiorano}, \citenamefont {Marinari}, \citenamefont
  {Martin-Mayor}, \citenamefont {Monforte-Garcia}, \citenamefont
  {Muñoz~Sudupe}, \citenamefont {Navarro}, \citenamefont {Parisi},
  \citenamefont {Perez-Gaviro}, \citenamefont {Ricci-Tersenghi}, \citenamefont
  {Ruiz-Lorenzo}, \citenamefont {Schifano}, \citenamefont {Seoane},
  \citenamefont {Taranc\'on}, \citenamefont {Tripiccione},\ and\ \citenamefont
  {Yllanes}}]{janus:16}%
  \BibitemOpen
  \bibfield  {author} {\bibinfo {author} {\bibfnamefont {M.}~\bibnamefont
  {Baity-Jesi}}, \bibinfo {author} {\bibfnamefont {E.}~\bibnamefont {Calore}},
  \bibinfo {author} {\bibfnamefont {A.}~\bibnamefont {Cruz}}, \bibinfo {author}
  {\bibfnamefont {L.~A.}\ \bibnamefont {Fernandez}}, \bibinfo {author}
  {\bibfnamefont {J.~M.}\ \bibnamefont {Gil-Narvi\'on}}, \bibinfo {author}
  {\bibfnamefont {A.}~\bibnamefont {Gordillo-Guerrero}}, \bibinfo {author}
  {\bibfnamefont {D.}~\bibnamefont {Iñiguez}}, \bibinfo {author}
  {\bibfnamefont {A.}~\bibnamefont {Maiorano}}, \bibinfo {author}
  {\bibfnamefont {E.}~\bibnamefont {Marinari}}, \bibinfo {author}
  {\bibfnamefont {V.}~\bibnamefont {Martin-Mayor}}, \bibinfo {author}
  {\bibfnamefont {J.}~\bibnamefont {Monforte-Garcia}}, \bibinfo {author}
  {\bibfnamefont {A.}~\bibnamefont {Muñoz~Sudupe}}, \bibinfo {author}
  {\bibfnamefont {D.}~\bibnamefont {Navarro}}, \bibinfo {author} {\bibfnamefont
  {G.}~\bibnamefont {Parisi}}, \bibinfo {author} {\bibfnamefont
  {S.}~\bibnamefont {Perez-Gaviro}}, \bibinfo {author} {\bibfnamefont
  {F.}~\bibnamefont {Ricci-Tersenghi}}, \bibinfo {author} {\bibfnamefont
  {J.~J.}\ \bibnamefont {Ruiz-Lorenzo}}, \bibinfo {author} {\bibfnamefont
  {S.~F.}\ \bibnamefont {Schifano}}, \bibinfo {author} {\bibfnamefont
  {B.}~\bibnamefont {Seoane}}, \bibinfo {author} {\bibfnamefont
  {A.}~\bibnamefont {Taranc\'on}}, \bibinfo {author} {\bibfnamefont
  {R.}~\bibnamefont {Tripiccione}}, \ and\ \bibinfo {author} {\bibfnamefont
  {D.}~\bibnamefont {Yllanes}},\ }\href {\doibase 10.1073/pnas.1621242114}
  {\bibfield  {journal} {\bibinfo  {journal} {Proceedings of the National
  Academy of Sciences}\ }\textbf {\bibinfo {volume} {114}},\ \bibinfo {pages}
  {1838} (\bibinfo {year} {2017}{\natexlab{b}})}\BibitemShut {NoStop}%
\bibitem [{Note3()}]{Note3}%
  \BibitemOpen
  \bibinfo {note} {For the sake of simplicity, we have neglected the correction
  of order $\xi ^{-\theta /2}$ in the $H^2$ term in Eq.~\protect \textup {\hbox
  {\mathsurround \z@ \protect \normalfont (\ignorespaces \ref
  {eq:prediction}\unskip \@@italiccorr )}}.}\BibitemShut {Stop}%
\bibitem [{\citenamefont {Djurberg}\ \emph {et~al.}(1995)\citenamefont
  {Djurberg}, \citenamefont {Mattsson},\ and\ \citenamefont
  {Nordblad}}]{djurberg:95}%
  \BibitemOpen
  \bibfield  {author} {\bibinfo {author} {\bibfnamefont {C.}~\bibnamefont
  {Djurberg}}, \bibinfo {author} {\bibfnamefont {J.}~\bibnamefont {Mattsson}},
  \ and\ \bibinfo {author} {\bibfnamefont {P.}~\bibnamefont {Nordblad}},\
  }\href {\doibase 10.1209/0295-5075/29/2/010} {\bibfield  {journal} {\bibinfo
  {journal} {Europhysics Letters ({EPL})}\ }\textbf {\bibinfo {volume} {29}},\
  \bibinfo {pages} {163} (\bibinfo {year} {1995})}\BibitemShut {NoStop}%
\bibitem [{Note4()}]{Note4}%
  \BibitemOpen
  \bibinfo {note} {Note that the scaling form of Eq.~\protect \textup {\hbox
  {\mathsurround \z@ \protect \normalfont (\ignorespaces \ref
  {eq:prediction}\unskip \@@italiccorr )}} is \protect \emph {necessary} to
  obtain a good collapse. To support this, we show in the Appendix
   that the collapse of Fig.~\ref{fig:scaling_law} deteriorates if one assumes different scaling
  behaviors.}\BibitemShut {Stop}%
\end{thebibliography}
\end{document}